\documentclass[10pt,journal,compsoc]{IEEEtran}

\newcommand{\ourtitle}{{{\sf Extracting Patterns of Urban Activity \\ from Geotagged Social Data}}}

\usepackage[utf8]{inputenc}
\usepackage{url}
\usepackage{microtype}
\usepackage{booktabs}
\usepackage{epsfig}
\usepackage{amssymb}
\usepackage{amsmath}
\usepackage{amsfonts}
\usepackage{subfigure}
\usepackage{float}
\usepackage{color}
\usepackage{graphicx}
\usepackage{dsfont}
\usepackage{enumitem}
\usepackage{algorithm}
\usepackage{algorithmic}
\usepackage[scanall]{psfrag}
\usepackage{ntheorem}
\usepackage{mathtools}
\usepackage{caption}
\usepackage{setspace}
\usepackage[nocompress]{cite}

\makeatletter
\newcommand\footnoteref[1]{\protected@xdef\@thefnmark{\ref{#1}}\@footnotemark}
\makeatother

\newtheorem{definition}{Definition}

\newtheorem{problem}{Problem}

\newtheorem*{proposition-non}{Proposition}

\newcommand{\spara}[1]{\smallskip\noindent{\bf{#1}}}
\newcommand{\mpara}[1]{\medskip\noindent{\bf{#1}}}

\DeclareMathOperator*{\argmax}{argmax}

\newcommand{\geoxpl}{{{\sf GeoExplore}}}

\newcommand{\livehoods}{{{\sf Livehoods}}}
\newcommand{\hoodsquare}{{{\sf Hoodsquare}}}

\newcommand{\twitter}{{{\tt Twitter}}}
\newcommand{\fsq}{{{\tt Foursquare}}}

\newcommand{\condsim}{{{\tt condsim}}}
\newcommand{\jointsim}{{{\tt jointsim}}}

\newcommand{\squishlist}{\begin{list}{$\bullet$}
  { \setlength{\itemsep}{0pt}
     \setlength{\parsep}{3pt}
     \setlength{\topsep}{3pt}
     \setlength{\partopsep}{0pt}
     \setlength{\leftmargin}{1.5em}
     \setlength{\labelwidth}{1em}
     \setlength{\labelsep}{0.5em} } }
\newcommand{\squishend}{
\end{list}  }

\newcommand{\projectwebsite}{\url{http://mmathioudakis.github.io/geotopics/}}
\usepackage{colortbl,xcolor}
\definecolor{lightorange}{HTML}{FF6E40}
\definecolor{lightgreen}{HTML}{00E676}
\newcommand{\cwin}[1]{{\cellcolor{lightgreen} #1}}
\newcommand{\cloose}[1]{{\cellcolor{lightorange} #1}}

\begin{document}
\title{
\ourtitle
}


\author{Emre Çelikten,
        Géraud Le Falher,
        and~Michael~Mathioudakis
\IEEEcompsocitemizethanks{\IEEEcompsocthanksitem Emre Çelikten and Michael
Mathioudakis are with the Department of Computer Science and the
Helsinki Institute for Information Technology, Aalto University,
Finland.\protect\\
E-mail: firstname.lastname@aalto.fi
\IEEEcompsocthanksitem Géraud Le Falher is with
Inria, Univ. Lille, CNRS UMR 9189 -- CRIStAL\protect\\
F-59000 Lille, France.\protect\\
E-mail: geraud.le-falher@inria.fr}
\thanks{}}

\IEEEtitleabstractindextext{%
\begin{abstract}
Data generated on location-based social networks provide rich information on
the whereabouts of urban dwellers. Specifically, such data reveal who
spends time where, when, and on what type of activity (e.g., shopping at a
mall, or dining at a restaurant). That information can, in turn, be used to
describe city regions in terms of activity that takes place therein. For
example, the data might reveal that citizens visit one region mainly for
shopping in the morning, while another for dining in the evening.
Furthermore, once such a description is available, one can ask more elaborate
questions: What are the features that distinguish one region from another
-- is it simply the type of venues they host or is it the visitors they
attract?
What regions are similar across cities?

In this paper, we attempt to answer these questions using
publicly shared Foursquare data. In contrast with previous work, our method
makes use of a probabilistic model with minimal assumptions about the
data and thus relieves us from having to make arbitrary decisions in our
analysis (e.g., regarding the granularity of discovered regions or the
importance of different features). We perform an empirical comparison
with previous work and discuss insights
obtained through our findings.
\end{abstract}
}
\maketitle


\section{Introduction}

Cities are massive and complex systems, the organisation
of which we often find difficult to grasp as individuals.
Those who live in cities get to know aspects of them
through personal experiences: from the cramped bar where we celebrate
the success of our favorite sports team
to the quiet café where we read a book on Sunday morning.
As our daily lives become more digitized, those personal experiences leave
digital traces, that we can analyse to understand better how we experience
our cities.

In this work, we analyze data from location-based social networks with the goal
to understand how different locations within a city are associated
with different kinds of activity -- and to seek similar patterns across
cities. To offer an example, we aim to automatically discover a
decomposition of a city into (potentially overlapping) regions,
such that one region is possibly associated, say, with shopping centers
that are active in the morning, while another is associated with
dining venues that are active in the evening.
We take a probabilistic approach to the task, so as to relieve
ourselves from having to make arbitrary decisions about crucial aspects of the
analysis -- e.g., the number of such regions or the granularity level of the
analysis. This probabilistic approach also provides a principled way to argue
about the importance of
different features for our analysis -- e.g., is the separation of regions
mostly due to the different categories of venues therein, or is it due to the
different visitors they attract?

Our work belongs to the growing field of Urban
Computing~\cite{UrbanComputingSurvey14} and shares its motivation.
First, as an ever-increasing number of people live in cities~\cite{unreport14},
understanding how cities are structured is becoming more crucial. Such
structure indeed affects the quality of life for citizens (e.g., how much time we
spend commuting), influences real-life decisions (e.g., where to rent an
apartment or how much to price a house), and might reflect or even enforce social
patterns (e.g. segregation of citizens in different regions).
Second, switching perspective from the city to the people, the increasing
amount of data produced by urban dwellers offer new opportunities in
understanding how citizens experience their cities. This understanding opens
possibilities to improve the citizens' enjoyment of cities. For instance, by
matching similar regions across cities, we could improve the relevance of
out-of-town recommendations for travelers.

The data we use were produced on \fsq, a popular
location-based social network, and provide
rich information about the offline activity of users.
Specifically, one of the main functionalities of the platform is
enabling its users to generate {\it check-ins} that
inform their friends of their whereabouts.
Each check-in contains information that reveals {\it who} (which user) spends
time {\it where} (at what location), {\it when} (what time of day, what day of
week), and doing {\it what} (according to the kind of venue: shopping
at a grocery store, dining at a restaurant, and so on).
The dataset consists of a total of {\bf 11.5 million \fsq\ checkins},
generated by users around the globe (Section \ref{sec:dataset}).

In the rest of the paper, we proceed as follows.
For each city in the dataset, we learn a probabilistic model for the
geographic distribution of venues across the city.
The trained models associate different regions of the city with
venues of different description. These venue descriptions are expressed in
terms of data features such as the venue category,
as well as the time and users of the related check-ins (Section \ref{sec:model}).
From a technical point of view, we employ a {\it sparse-modeling}
approach~\cite{eisenstein2011sparse}, essentially enforcing that
a region will be associated with a distinct description only if that
is strongly supported by data.

Once such a model is learnt for each city in our dataset through an
expectation--maximization
algorithm (Section \ref{ssec:learning}), we examine how features are spatially
distributed within a city, illustrating the insights it provides for
some of the cities in our dataset (Section \ref{sec:features}).
Subsequently, we make use of the learnt models and address two tasks.
\begin{itemize}
    \item The first is to understand which features among the ones we consider
  are more significant for distinguishing regions in the same city
  (Section \ref{sec:neighborhood_description}).
  Somewhat surprisingly, we find that
  {\it who} visits a venue has higher distinguishing power than
  other features (e.g., the category of the venue). This is a finding
  that is consistent across the cities that we trained a model for.
    \item The second is to find similar regions across different cities
    (Section \ref{sec:similar}). To quantify the similarity of two regions, we
  define two measures that have a natural interpretation within the
  probabilistic framework of this work. First, we discuss the properties
  of each measure, in what cases one would employ each, and describe
  how one would employ them in an algorithmic search for similar regions
  across cities.
  Subsequently, we employ them on our dataset and find that the regions
  automatically detected in our model provide very well matching regions
\end{itemize}

Having provided the results of our analysis, we compare our modeling
approach to previously used approaches. Our empirical evaluation in
Section \ref{sec:other_approaches} shows that our
approach outperforms previous attempts~\cite{Livehoods12,Hoodsquare13}
in terms of predictive performance as well as finding more distinctly described
regions.

Finally, we review related work (Section \ref{sec:related}),
and discuss possible extensions and improvements of our
own work in Section \ref{sec:conclusions}.
Our code and anonymized versions of our dataset will be made publicly
available at \projectwebsite{} upon publication.

\section{Data \& Model}
\label{sec:datamodel}

\subsection{Dataset}
\label{sec:dataset}

Our dataset consists of geo-tagged activity from \fsq{},
a popular location-based social network that, as of 2015, claims
more than $50$ million users\footnote{\url{https://foursquare.com/about/}.}.
It enables users to share their current location with friends, rate
and review venues they visit, and read reviews of other users.
\fsq\ users share their activity by generating
{\it check-ins} using a dedicated mobile application\footnote{The Swarm
application, \url{http://www.swarmapp.com}.}.
Each check-in is associated with a web page that contains information
about the user, the venue, and other details of the visit.
Each venue is also associated with a public web page that contains
information about it --- notably the city it belongs to, its geographic
coordinates and a category, such as \emph{Food} or
\emph{Nightlife Spot}.

According to \fsq's policy, check-ins are private by default,
i.e., they become publicly accessible only at the users' discretion.
This is the case, for example, when users opt to share their check-ins
publicly via \twitter\footnote{\url{http://twitter.com}}, a popular micro-blogging
platform. We were thus able to obtain \fsq\ data by retrieving check-ins shared on
\twitter{} during the summer 2015. In order to have data for the whole year, we
add data from a previous work~\cite{Thesis15} collected in the same way.
We did not apply any filtering during data collection,
but for the purposes of this work, we focus on the 40 cities with highest
volume of check-ins in our data. The code for collecting the
data is made publicly available on
{\it github}\footnote{\projectwebsite}.

The data from the 40 selected cities consist of
approximately 6.7 million check-ins with 498 thousand unique users, in total.
As a post-processing step,
we removed the check-ins of users who contributed check-ins at less than five
different venues,
resulting in 6.3 million check-ins and approximately 284 thousand unique
users (i.e. more than 7,000 unique users per city) in our working dataset.
Details of the dataset can be found in Table~\ref{tab:dataset}.


Although our dataset is made of three type of entities (users, checkins, and
venues), we take a venue-centric view of the data. Indeed, venues are the
entities associated with the largest amount of information and furthermore, we
want to analyse how venues are distributed geographically within a city.
Specifically, we associate the following information with each venue.
\squishlist
\item A geographic {\tt location}, expressed as a longitude-latitude pair
    of geographic coordinates.
\item The {\tt category} of the venue, as specified by \fsq's
    taxonomy (e.g., `Art Gallery', `Irani Cafe', `Mini Golf').
    If more than one categories are associated with one venue, we keep
    the one that is designated as the `main category'.
\item A list of all check-ins associated with this venue
    in the working dataset. Each check-in is a triplet that contains the
    following data:
\squishlist
  \item The unique identifier of the {\tt user} who performed it;
  \item The {\tt day of the week} when the check-in occurred, expressed as
  a categorical variable with values {\it Monday}, {\it Tuesday}, ...,
  {\it Sunday};
  \item The {\tt time of the day} when the check-in occurred, expressed as a
  categorical variable with values \textit{morning}, \textit{noon}, ...,
  \textit{late night}, defined according to Table~\ref{table:timeoftheday}.
\squishend
\squishend

According to this view, each venue is a single data point described in terms of
five {\bf features} -- namely {\tt location}, {\tt category}, {\tt users},
{\tt times of day} and {\tt days of week}, and it takes a list of values for
each feature.
For the first two -- {\tt location} and
{\tt category} -- the size of the list is always $1$ -- i.e., each venue is
associated with a single location and a single category.
Moreover, {\tt location} values are {\it continuous} two-dimensional,
while for all other features the values are {\it categorical}.
For the categorical features, i.e., {\tt category},
{\tt times of day}, {\tt days of week}, and {\tt users},
we'll be using the term
{\it dimensionality} to refer to the number of values they can take.
For example, the dimensionality of {\tt times of day} is always $6$,
that of {\tt days of week} $7$, that of {\tt category} is about $700$,
and that of {\tt users} has an average
of more than $10,000$ within each city in the dataset.

{\setlength{\tabcolsep}{2pt}
    \begin{table}[pht]
    \caption{{\tt Time of the Day} intervals}
      \begin{center}
          \begin{tabular}{|r|c|c|c|c|c|c|}
            \hline
            & \textit{morning} & \textit{noon} & \textit{afternoon}
            & \textit{evening} & \textit{night} & \textit{late night} \\ \hline
            from &  6 am & 10 am & 2 pm  & 6 pm   & 10 pm  & 2 am  \\
            to   & 10 am &  2 pm & 6 pm  & 10 pm  &  2 am  & 6 am \\ \hline
          \end{tabular}
      \end{center}
      \label{table:timeoftheday}
    \end{table}
}

\begin{table}[t]
  \centering
  \setlength{\tabcolsep}{2.5pt}
  \footnotesize
  \caption{Number of check-ins and venues for the 18 (out of 40) cities
  with most data.
  They cover a large part of the world (Americas, Europe,
  Middle East and Asia).
\label{tab:dataset}}
\begin{tabular}{crrcrr}
  \toprule
  City         & Check-ins & Venues & City           & Check-ins & Venues \\
  \midrule
  Ankara       & 104,002   & 16,983 & Mexico City    & 122,561   & 28,779 \\
  Barcelona    & 213,859   & 20,353 & Moscow         & 397,008   & 51,871 \\
  Berlin       & 141,161   & 18,544 & New York       & 1,007,377 & 75,721 \\
  Chicago      & 306,296   & 27,949 & Paris          & 284,776   & 28,489 \\
  Istanbul     & 578,042   & 69,008 & Rio de Janeiro & 47,743    & 13,394 \\
  Izmir        & 190,303   & 20,529 & San Francisco  & 432,625   & 22,384 \\
  Kuala Lumpur & 147,103   & 22,594 & Seattle        & 103,575   & 10,591 \\
  London       & 234,744   & 26,453 & Washington     & 412,863   & 20,122 \\
  Los Angeles  & 367,624   & 36,086 & Tokyo          & 214,493   & 38,117 \\
  \midrule
  \multicolumn{6}{c}{For the 40 cities; 6,335,350 Check-ins and 749,097 Venues in total}  \\
  \bottomrule
\end{tabular}
\end{table}

\subsection{Model Definition}
\label{sec:model}

Our analysis is based on a generative model that describes the venues
we observe in a city. More precisely,
each data point generated by the model corresponds to a single venue,
and is associated with a list of values for each {\it feature}
described in Section~\ref{sec:dataset}.

Remember that our goal is to uncover associations between geographic
locations and other features of venues. Such associations
are captured as $k$ {\it topics} in the model --
i.e., each data point is assigned probabilistically to
one topic and different topics generate data venues with different
distributions of features.
As an example, one topic might generate venues (data points) that are located in
the south of a city (feature: {\tt location}) and are particularly popular
in the morning (feature: {\tt time of the day}), while
another might generate venues that are located in the north of a city
(feature: {\tt location}) and predominantly restaurants, bars, and night-clubs
(feature: {\tt category}).

\smallskip
Specifically, to generate one data point, the model performs the following
steps:
\squishlist
\item Select one ($1$) out of $k$ available topics $\{1, 2, \ldots, k\}$
according to a multinomial probability distribution
$\theta = (\theta_1, \theta_2, \ldots, \theta_k)$.
Let the selected topic be $z$.
\item Generate a geographic location $loc = (x,y)$ from a bivariate Gaussian
 distribution
with center $c = c_z$ and variance matrix $\Sigma = \Sigma_z$.
\item For the $i$-th categorical feature,
generate a list $\mathbf{u} = \mathbf{u_i}$ of $N = N_i$ items, where $N_i$
is specified as input for this data point. Each element in the list is selected
randomly with replacement from a set $U = U_i = \{u_1, u_2, \ldots, u_m\}$
according to multinomial probability
$\beta = \beta_z^{(i)} = \ (\beta_{z}^{i}\,_{|1}, \beta_{z}^{i}\,_{|2}, \ldots,
\beta_{z}^{i}\,_{|m})$,
with $\beta_{z}^{i}\,_{|j} \geq 0$ and
$\sum_{j = 1 .. m} \beta_{z}^{i}\,_{|j} = 1$.
\squishend
\smallskip

\begin{figure}[t]
  \includegraphics[width=0.90\columnwidth]{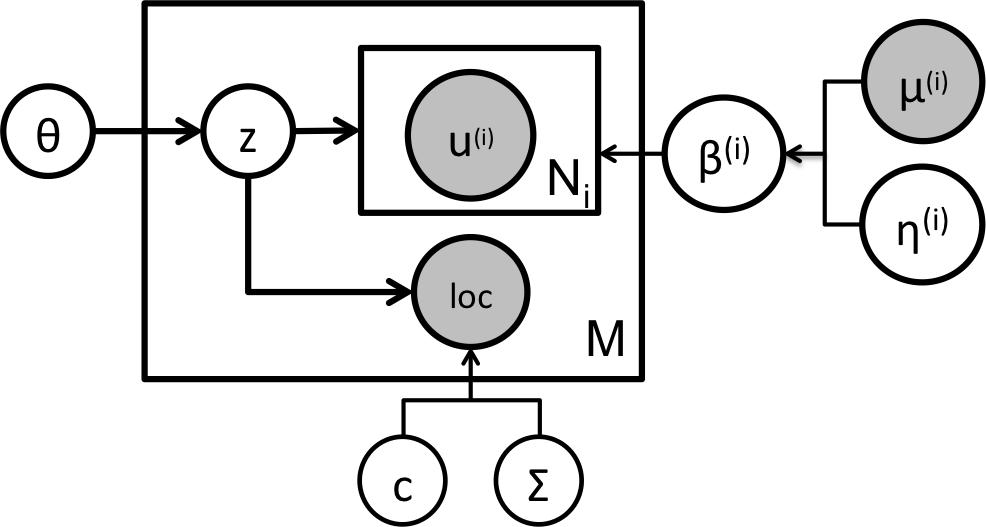}
  \caption{Generative Model. Note that only the $i$-th categorical feature
  is depicted.}
  \label{fig:plate-basic}
\end{figure}

The model is depicted in the plate diagram of
Figure~\ref{fig:plate-basic}. We stress that there is a different multinomial
distribution $\beta = \beta_z^{(i)}$ for the $z$-th topic and $i$-th feature.
We will be using non-subscript notation ($\beta$ instead of $\beta_z^{(i)}$)
when we might refer to any such distribution vector -- and do the same with
other notation symbols.

The procedure described above is repeated $M$ times to generate a
dataset of size $M$ (i.e., $M$ venues).

Moreover, we assume that, for the $i$-th feature, the probability distribution
$\beta_z^i$ is derived from a probability distribution $\mu^i$
and a deviation vector $\eta = \eta_z^i$ according to the following
formula:
\begin{equation}
  \beta_z^i \propto \exp(\mu^{i} + \eta_{z}^{i}).
  \label{eq:deviation}
\end{equation}
Firstly, the distribution
$\mu$ models the `global' log-probability
that the model generate an element $u\in U$.
The model makes the assumption that all such probability distributions $\mu^i$
are equally likely (uniform prior).
Secondly, the deviation vector
$\eta_z^i$ quantifies how much the distribution $\beta_z$ of topic
$z$ deviates from global distribution $\mu^i$.
The model makes the assumption that each value of vector $\eta$ is
selected at random with prior probability
\begin{equation}
  \log p(\eta_z^{(i)}) = - \lambda\cdot |\eta_z^{(i)}| + \text{constant}
\end{equation}
for some coefficient $\lambda$, provided as input.
The model thus penalizes large deviations from `global' vectors $\mu$,
thus leading to `sparse' vectors $\eta_z^i$.
The motivation for favoring sparce vectors $\eta$ is that we wish
to associate different topics with different distributions $\beta$ only if
we have significant support from the data.
For the remaining parameters of the model, we make the following assumptions:
all centers $c_z$ are equally likely (uniform prior),
the value of $\Sigma_z$ has a Jeffreys prior,
\begin{equation}
\log p(\Sigma_z) = - \log \|\Sigma_z\| + \text{constant}
\end{equation}
and all $\theta$ vectors are equally likely
(uniform prior).

\subsection{Learning}
\label{ssec:learning}

We learn one instance $I$ of the model for each city in our dataset.
Formally, learning corresponds to the optimization problem below.
\begin{problem}
Given a set of data points $D = \{d_1, d_2, \ldots, d_{_M}\}$,
and the generative model $I$,
find global log-probabilities $\mu$,
topic distribution $\theta$,
deviation vectors $\eta$, bivariate Gaussian centers $c$ and
covariances $\Sigma$ that maximize the probability
\begin{equation*}
L = p(\{\mu\}, \theta, \{\eta\}, \{c\}, \{\Sigma\} \; | \; D, I).
\end{equation*}
\label{problem:basic-problem}
\end{problem}

We perform the optimization by partitioning the dataset into a training and
test dataset and following a standard validation procedure.
During training, we keep $k$ and $\lambda$ fixed and optimize the remaining
parameters of the model on the training dataset ($80\%$ of all data points).
We then evaluate the performance of the model on the test dataset
($20\%$ of all data points), by calculating the log-likelihood of the test
data under the model produced during training.
We repeat the procedure for a range of values for $k$ and $\lambda$ to
select an optimal configuration.

A max-likelihood vector $\mu$ is computed once for each feature
from the raw relative frequencies of observed values of that feature in the
dataset. For fixed $k$ and $\lambda$, the maximum-likelihood value of the remaining
parameters can be computed with a standard expectation--maximization algorithm.
The steps of the
algorithms are provided below,
\spara

\mpara{\underline{E-Step}}
\begin{eqnarray*}
\log{q_d(z)} & := & \sum_{i=1..m}n_d^{(i)}\log{\beta_z^{(i)}}
     + \log{\mathcal{N}(loc_d;c_z,\Sigma_z)} + \\
     &  & + \log{\theta_z} + constant;\; \sum_z q_d(z) = 1
\end{eqnarray*}
\spara{\underline{M-Step}}
\begin{eqnarray*}
\theta_z & := & \sum_{d\in D}{q_d(z)},\;\text{normalized to}\;\sum_z \theta_z = 1 \\
c_z & = & \frac{\sum_d q_d(z) \cdot loc_d}{\sum_d q_d(z)} \\
\Sigma_z  & := & \dfrac{\sum_{d\in D}{q_d(z)
(loc_d - c_z)(loc_d - c_z)^T}}{\sum_{d\in D}{q_d(z)} + 4} \\
\{\eta_z^{(i)}\} & := & \argmax_{\{\eta_z^{(i)}\}} \sum_{d\in D} q_d(z)\cdot n_d^{(i)} \cdot \log\beta_z^{(i)} - \lambda \cdot |\eta_z^{(i)}|,
\end{eqnarray*}
with $n_d^{(i)}$ the number of times the $i$-th element appears in data point $d$,
and the latter optimization (for $\eta$ values) performed numerically.

\begin{figure}[tb]
  \centering
        \includegraphics[width=0.95\linewidth]{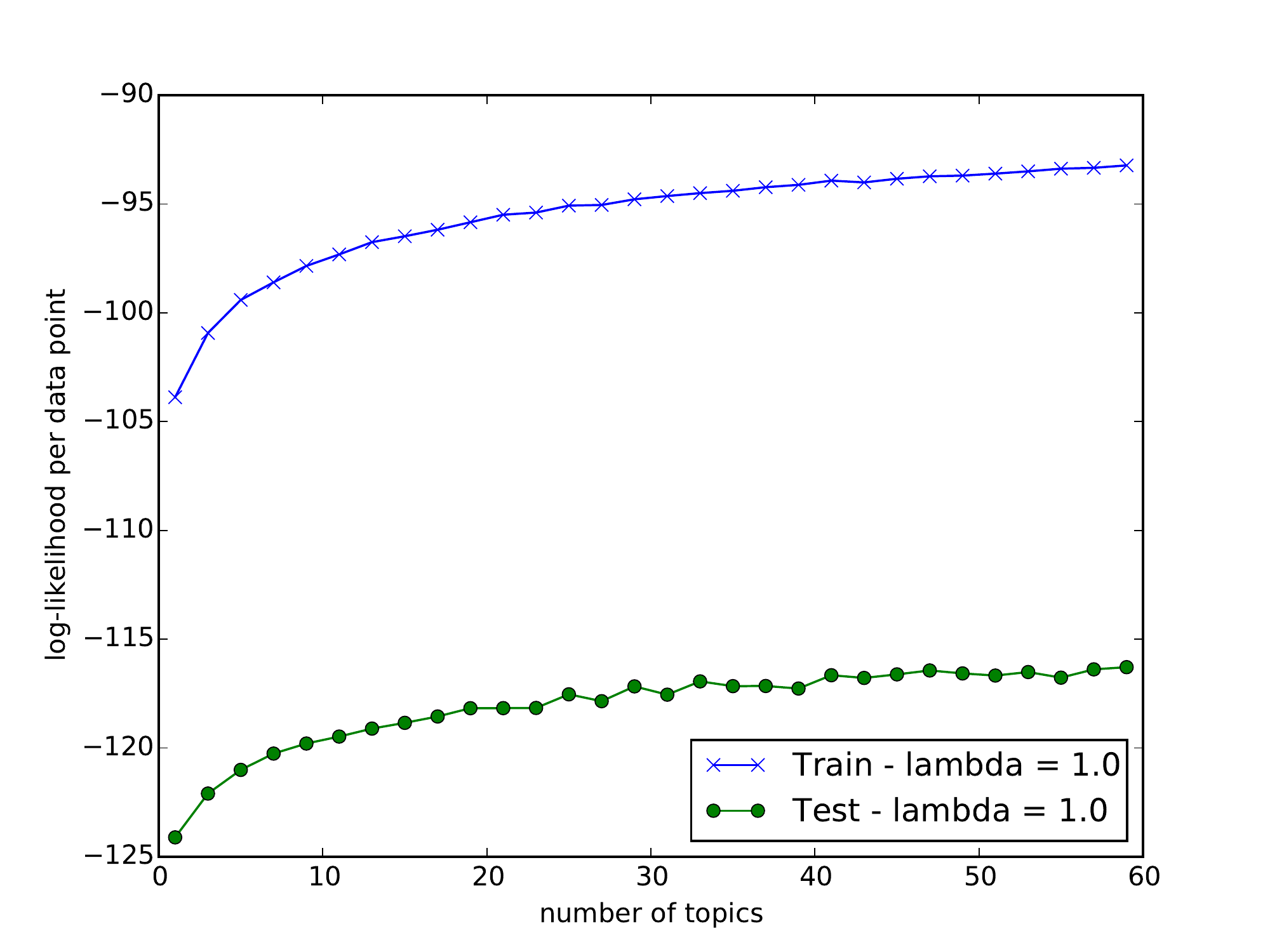}
  \caption{Log-likelihood per data point on the training and test datasets
  of Paris, for $\lambda = 1$ and increasing $k$.
        \label{fig:paris-logl}}
\end{figure}

To optimize with respect to $k$ and $\lambda$, we experiment with a grid
of values and
select the pair of values with the best performance on the test set.
We found that $\lambda \approx 1$ worked well for
all cities we experimented with, while improvement reached a plateau for
values of $k$ near $k \approx 50-55$. Figure~\ref{fig:paris-logl}
shows the training plots for the city of Paris; similar
patterns are observed for the other cities in the dataset.

\subsection*{Practical issues}

In training, we came across main memory capacity issues during the estimation of
model variables related to feature {\tt users} -- particularly for the
corresponding multinomial deviation vector $\eta$.
We believe that was due to the large dimensionality of the {\tt users}
feature. Indeed, on average, our data contain activity
of more than $10,000$ users per city -- while the dimensionality of
features {\tt time of the day}, {\tt day of the week}, {\tt category} is
much lower $6$, $7$, and $\approx 400$, respectively.

To deal with the issue, we use SVD to reduce the dimensionality of feature
{\tt users}. Our goal is to
partition the users $\{ u_1, u_2, \ldots, u_{_M} \}$ that appear in one city
into $d$ groups $U_1, U_2, \ldots, U_{d}$, in such a way that
$U_i$ groups together users that check-in at the same venues.
SVD captures nicely such semantics and this property has been used in many
settings (e.g., latent semantic indexing~\cite{deerwester1990indexing}).
Once the partition is produced, we treat all users in group $U_i$,
$i = 1..d$, as the same user (a `super-user'), thus reducing the dimensionality
of feature {\tt users} to $d$.

Specifically, for each city, we
consider the users-venue matrix $M$. The $(i, j)$ entry of
matrix $M$ contains the number of check-ins observed for the $i$-th user at the
the $j$-th venue in the city. Subsequently, we use SVD to compute the
$d$ right-eigenvectors $\mathbf{v_1}, \mathbf{v_2}, \ldots, \mathbf{v_d}$ of
$M$. Note that the dimensionality of each such eigenvector $\mathbf{v}$
is equal to the dimensionality of the {\tt users} feature for a given city.
Finally, we partition the users into $d$ groups:
we assign the $i$-th user to group $g \in \{ 0, 1, \ldots, d \}$
such that
\begin{equation*}
g = \argmax_{j\in \{1, 2, \ldots, d\}} \{ \mathbf{v_1}[i],
\mathbf{v_2}[i], \ldots, \mathbf{v_d}[i] \}
\end{equation*}
This provides naturally the partition $U_1, U_2, \ldots, U_{d}$
we aimed to identify.

\section{Likely and distinctive feature values}
\label{sec:features}
Following the learning procedure defined in Section~\ref{sec:datamodel},
we learn a single model for each city in our data.
The value of these model instances is that they offer a principled way
to answer quesions that we cannot answer from raw data alone.
To provide an example, suppose that an acquaintenance from abroad visits our
city and asks ``if I stay at location $l$ of the city, what is the most
common venue category I find there?". Raw data do not provide an
immediate answer to the question. They do allow us, for example, to provide
answers of the form ``within radius $r$ from $l$, the most common venue
category is {\it c} with $n$ out of $N$ venues''. However such answers
would depend on quantity $r$, that was not provided as input -- and
would probably never be, if our friend does not have any knowledge
about the city. Selecting too small a value for $r$ (e.g., a few meters),
would make the answer sensitive to the exact location $l$; selecting
too large a value for $r$ (e.g., a few kilometers), would make the answer
insensitive to the exact location $l$.

The unsupervised learning approach
we take allows us to avoid such arbitrary choices in a principled manner.
It learns topics associated with gaussian distributions as regions, whose
size is  learned from the data; and under a model instance $I$, it allows
us to answer our friend's question by simply considering the probability
$$ p(\text{category} = c| \text{location} = l; I)$$
that at the given location we find a venue of category $c$, and
answering with the category that is associated with the highest
probability value.

Given such model
instances, we explore the geographical distribution of
venues for the corresponding cities.
Due to space constraints, we provide only a few examples here and
will provide a complete list of findings from this section on the project's
webpage\footnote{\projectwebsite{}}.

\begin{figure*}[tb]
  \centering
    \subfigure[category]{
        \includegraphics[height=3.26in]{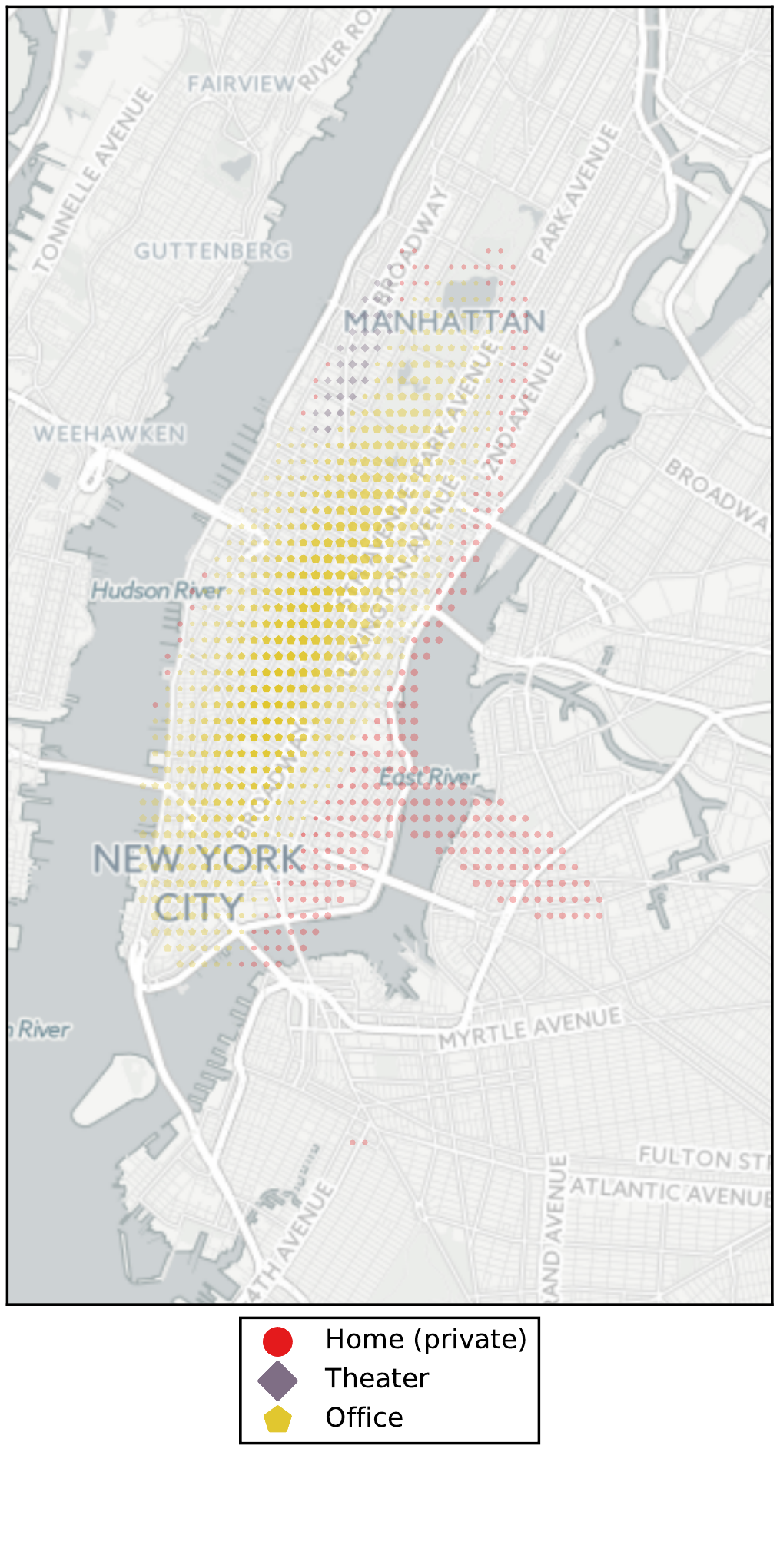}
        \label{fig:newyork_category_likely}
    }
    \subfigure[time of day]{
        \includegraphics[height=3.26in]{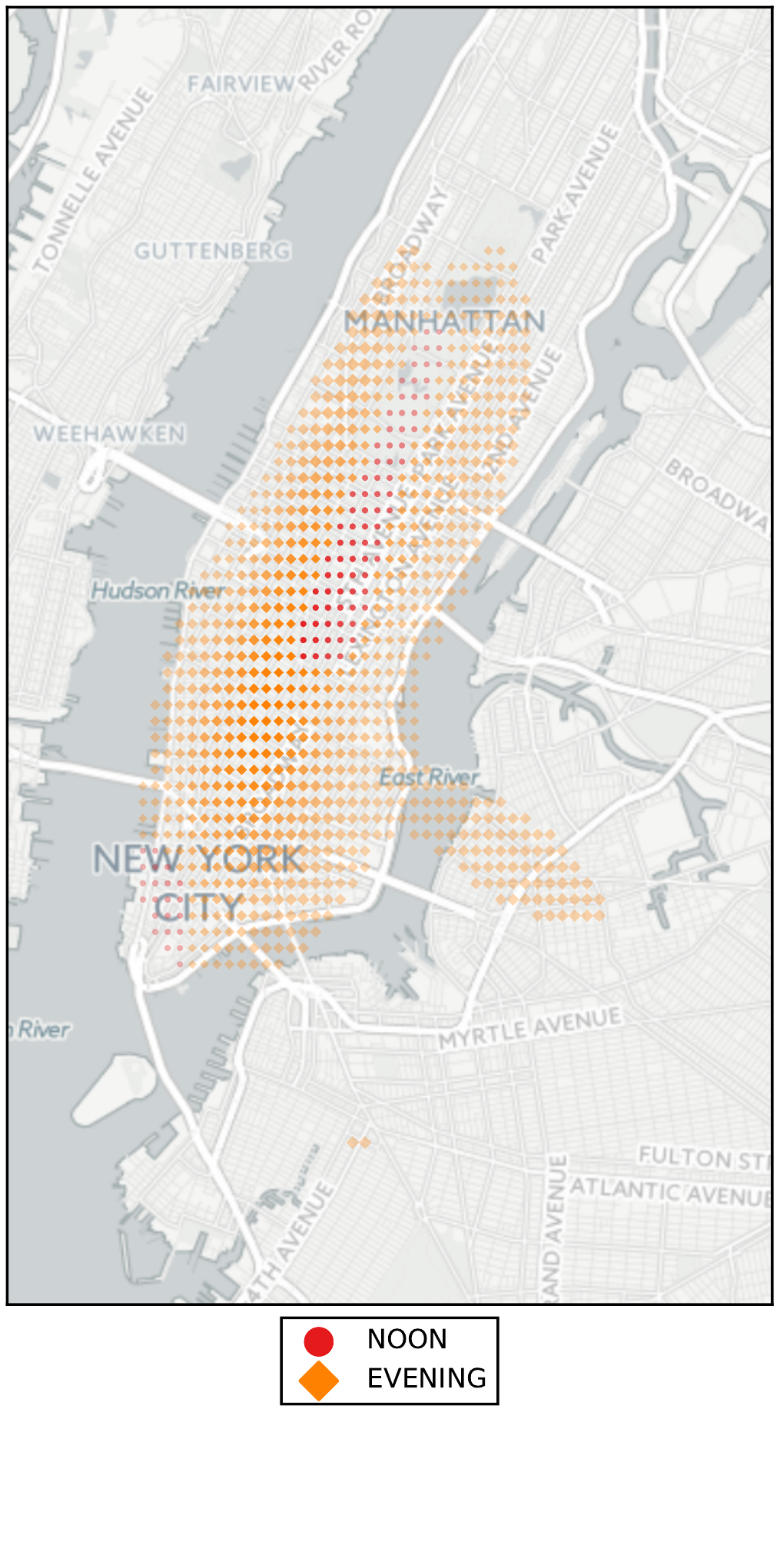}
        \label{fig:newyork_timeofday_likely}
    }
    \subfigure[day of week]{
        \includegraphics[height=3.26in]{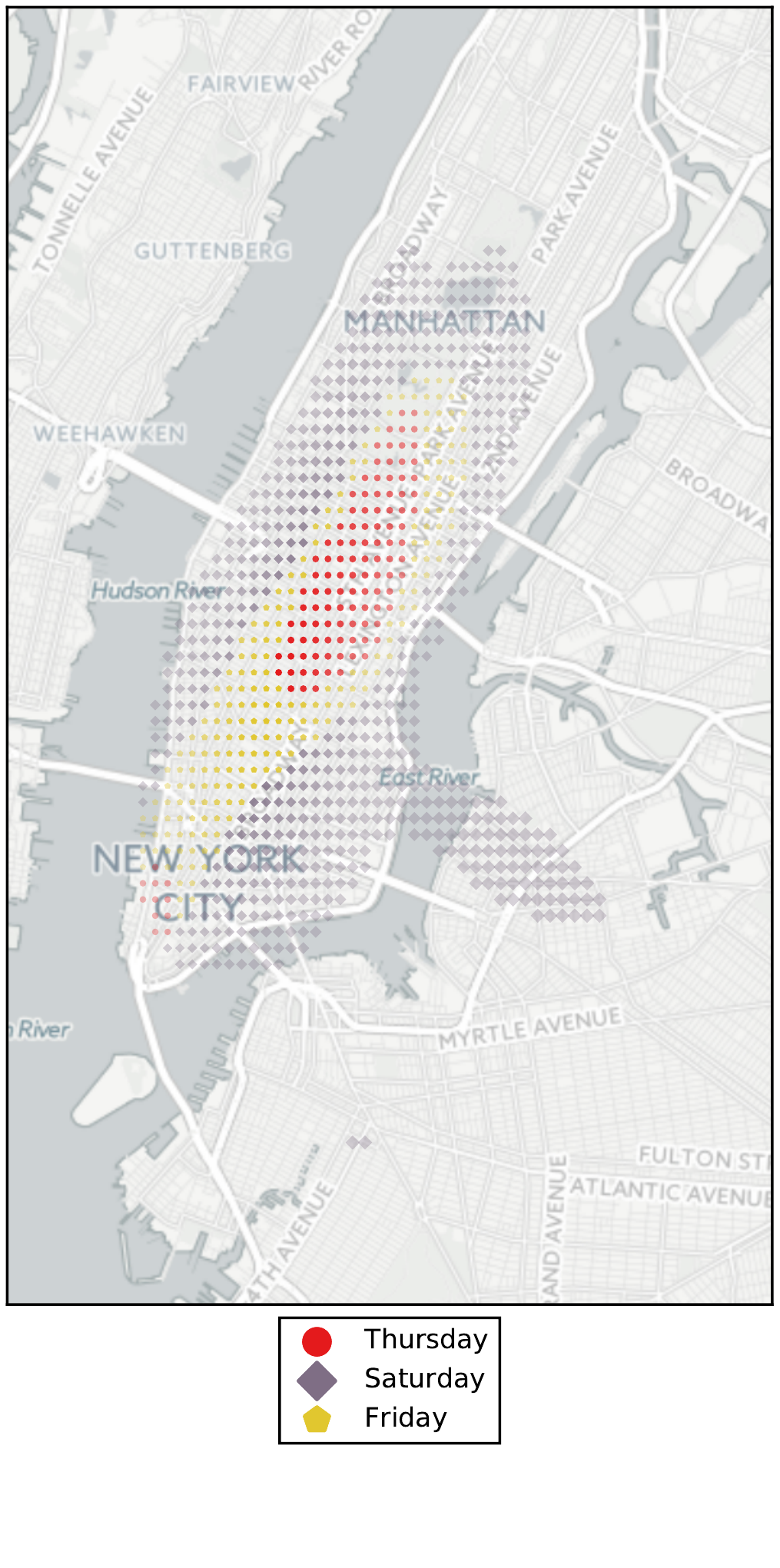}
        \label{fig:newyork_dayofweek_likely}
    }
    \caption{Most likely {\tt category} and checkin {\tt time of day, day of week}
    across Manhattan. Note that the transparency of each point is equal
    to the probability that a venue is located at that point.}
\end{figure*}

\spara{Most likely feature value} Suppose a venue is placed at a given location
$loc = (x, y)$ -- what is the category most likely associated with it?
In other words, we are asking for the category that maximizes the expression
\begin{equation}
p(\text{category} = c \; | \; \text{location} = l; I)
\end{equation}
that we just discussed above.
We use our model to answer this question for {\it New York}.
The results are shown in Figure~\ref{fig:newyork_category_likely}.
We can ask a similar question for the remaining categorical features
represented in our model. For example, suppose a venue is placed at a given
location, what is the most likely time a check-in occurs at that venue?
\footnote{Note that the question is conditioned on both a venue with known
location, not the location alone.}
The results for {\it New York} are given in
Figure~\ref{fig:newyork_timeofday_likely}.

\spara{Most distinctive feature value} Looking again at
Figure~\ref{fig:newyork_timeofday_likely}, we see that {\it evening}
check-ins dominate the map: for many locations in Manhattan, a venue placed there
is most likely to receive a check-in during the evening.
One simple explanation for this is that overwhelmingly many check-ins in our
data for this city  occur in the evening, as we see in
table~\ref{table:nycheckins}.

{\setlength{\tabcolsep}{2pt}
\begin{table}[h!]
\caption{New York City check-ins in thousands}
\begin{center}
\begin{tabular}{|c|c|c|c|c|c|}
 \hline
 morning & noon    & afternoon & evening & night   & latenight \\ \hline
106  & 219 & 240   & 333 & 118 & 25    \\ \hline
\end{tabular}
\end{center}
\label{table:nycheckins}
\end{table}
}

\begin{figure*}[tb]
  \centering
  \subfigure[category]{
    \includegraphics[height=3.26in]{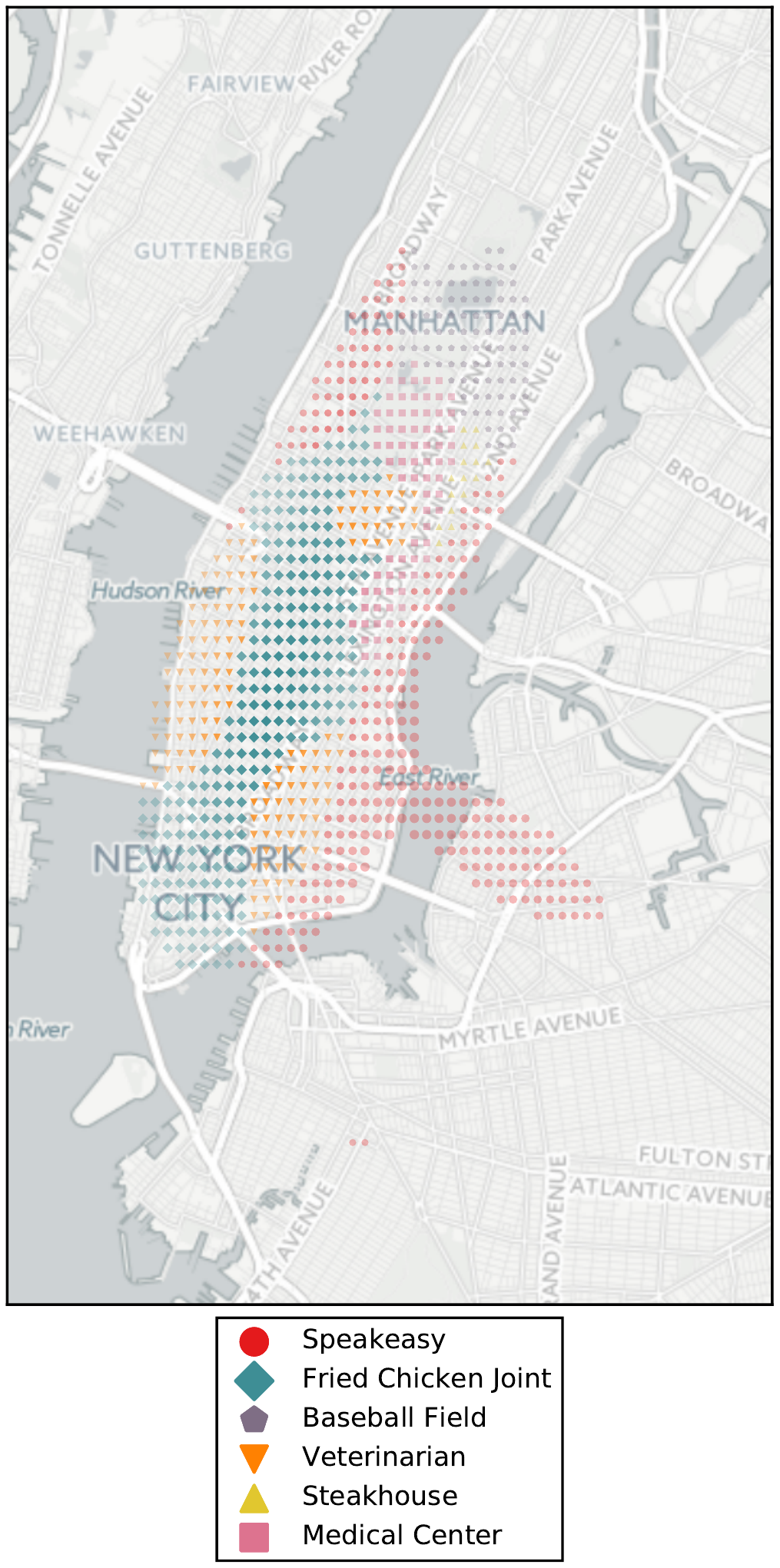}
    \label{fig:newyork_category_distinctive}
  }
  \subfigure[time of day]{
    \includegraphics[height=3.26in]{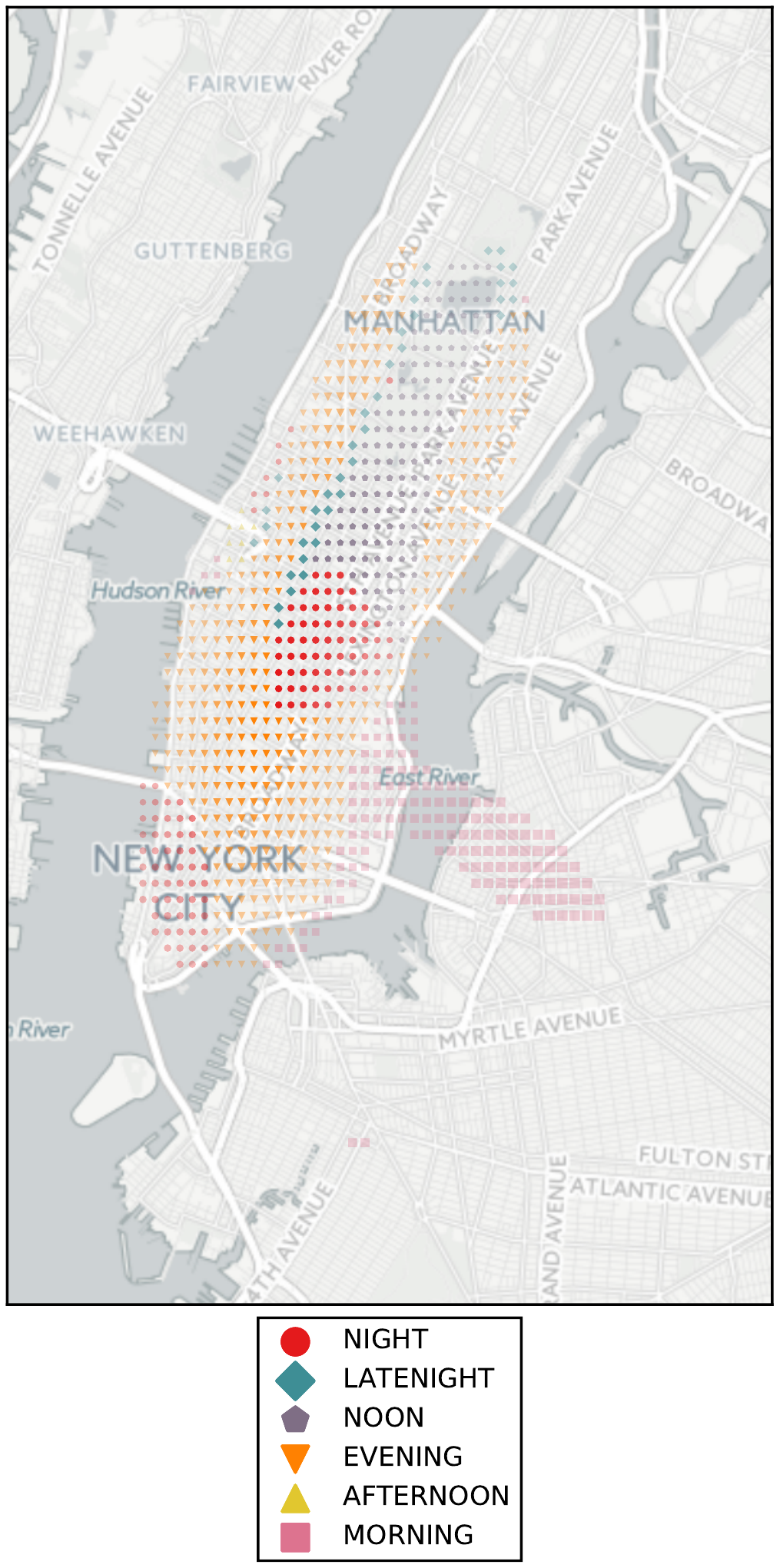}
    \label{fig:newyork_timeofday_distinctive}
  }
  \subfigure[day of week]{
      \includegraphics[height=3.26in]{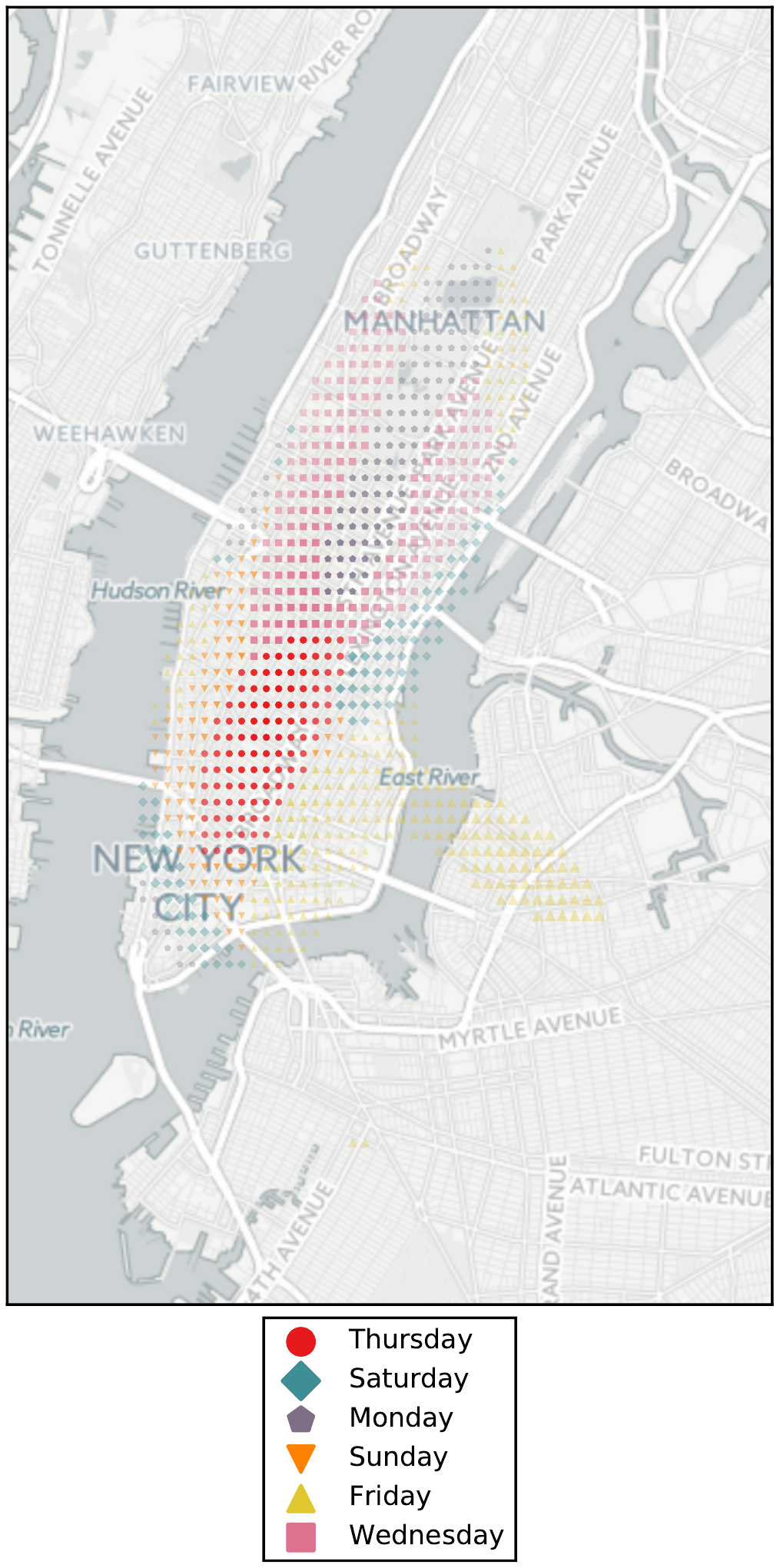}
      \label{fig:newyork_dayofweek_distinctive}
  }
  \caption{Most distinctive {\tt category} and checkin {\tt time of day, day of week}
  across Manhattan. The transparency of each point is equal
  to the probability that a venue is located at that point.}
\end{figure*}

Nevertheless, some areas of the city are more highly associated with morning
check-ins than others.
In formal terms, for a given location, let us consider the ratio
of the probability that the {\tt time of day} a check-in occurs takes a
particular value (`morning', `noon', etc) over the probability that a check-in
takes that value over the entire city. Arguably, that ratio expresses
how distinctive that value is for this particular location.
Formally, it is expressed as follows.
\begin{equation}
  \frac{p(\text{category} = c \; | \; \text{location} = l; I)}
    {p(\text{category} = c \; | \; I}
\end{equation}
For example,
suppose that a venue at a particular location $loc$
receives a check-in in the morning with probability $30$\%;
and that on average across the city venues,
a venue would receive a check-in in the morning with probability only $1\%$.
Then, we can say that location $loc$ is associated with venues that are
distinguished for the relatively high frequency of morning check-ins.
Figure~\ref{fig:newyork_timeofday_distinctive}
indicates the most distinctive
time of check-in across New York. We can ask a similar question for other
categorical features.  For example, {\it what is the most distinctive
category for the same location?} The results for {\it New York}
are shown in Figure~\ref{fig:newyork_category_distinctive}.

\section{Feature analysis}
\label{sec:neighborhood_description}

In the previous section, we employed the model instance of a city to
ask questions about the geographical distribution of a given feature.
In this section, we study the importance
of each feature in distinguishing the different topics that
define a model instance. To further illuminate the question, let us remember
that the model is built upon a set of features $\{ X \}$ and that the
distribution of each feature is allowed to vary across topics.
In plain terms, the question we ask is the following: {\it if we were forced
to fix the distribution of feature $X$ across topics, how much would that hurt
the predictive performance of the model?} This question is important, not
just for the purposes of feature selection in case one wanted to employ a
simpler model, but also because it allows us to suggest on what features
future work should be focused, in order to understand better urban
activity.

To be more specific, let us first consider the categorical features
in our model: {\tt category}, {\tt users}, {\tt time of the day},
{\tt day of the week}. Within each topic of a model instance,
the distribution $\beta$ of each of the aforementioned features
deviates from the overall distribution $\mu$ by a vector
$\eta$ (Equation~\ref{eq:deviation}).
To quantify to which extent a single categorical feature $X$ contributes to
the variance across topics, we perform a simple ablation study. That is,
we select the same training set as for our best model, keep the value of
parameters $k$ and $\lambda$, and train a model instance by fixing
the $\eta$ of feature $X$ equal to zero.
$$\text{ for categorical feature} X:\; \text{set}\; \eta = 0$$
Subsequently we compare the
log likelihood of both models (the best one and the ablated one) over all
the data for the given city and measure the log-likelihood drop between
the two. The higher this drop, the higher the importance of that
feature in explaining the variance across topics.

We perform a similar procedure for the {\tt location} feature.
Specifically, for the ablated model, we replace the bivariate Gaussian
distribution $G_z$ associated with each model with
a distribution $G_0$ that remains fixed across topics. $G_0$ is set
to be the mixture of Gaussians $G_z$ across topics $z$, with
mixture proportions equal to topic proportions $\theta_z$.
$$\text{ for
{\tt location}}:\;\text{set}\; G_0 = \text{mixture} \{G_z, \theta_z\}$$

\begin{figure}
  \centering
  \includegraphics[width=0.48\textwidth]{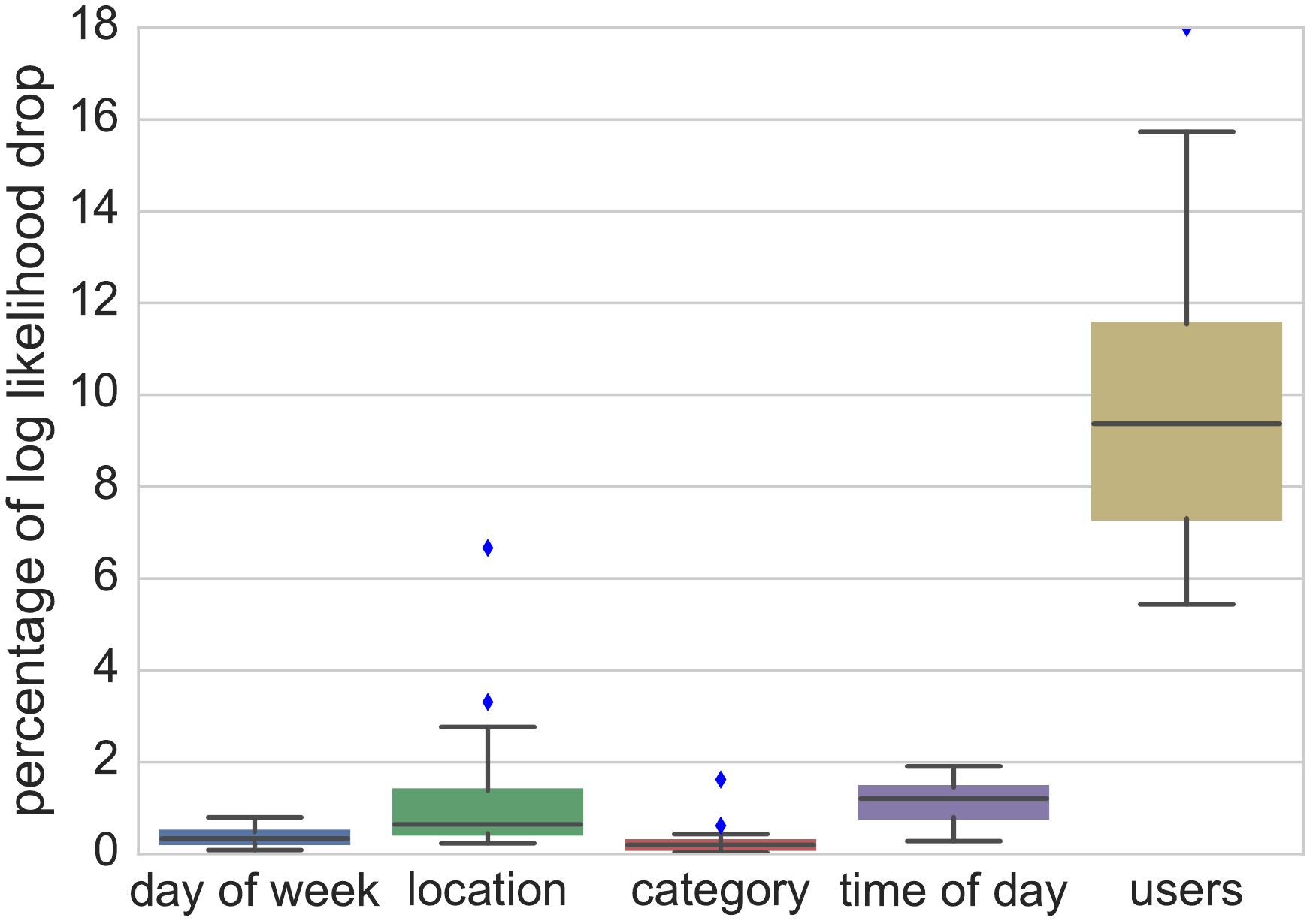}
  \caption{Contribution of each feature to the data likelihood. The boxplots
    summarize how much the log likelihood drops once we fix the distribution
    of a single feature across topics. We observe a consistent
    behavior across cities, in that the variance of {\tt users} across
    topics is most important for the predictive performance of the model.}
  \label{fig:feature_contributions}
\end{figure}

Results are summarized for all cities in Figure \ref{fig:feature_contributions}.
The immediate observation is that {\tt users} prominently stand out as
a feature and that this is consistent across all cities. This suggests that, at least for the
urban activity represented in our dataset, {\bf who} visits a venue
has a more important role to play in distinguishing different venues,
than where the venues are located and when they are active.
Among the remaining features, {\tt location} and {\tt time of the day},
are consistently more important across cities than {\tt day of the week}
and {\tt category}.

\section{Similar regions across cities}
\label{sec:similar}

In this section, we address the task of discovering similar regions
across different cities.
Addressing this task would be useful, for example, to generate touristic
recommendations for people who visit a foreign city or generally
to develop a better understanding of a foreign city based on knowledge
from one's own city.
Specifically, we are given the
trained models of two (2) cities as input and aim to identify one region
from each city so that a similarity measure for the two regions is maximized.
Following the conventions of this work, each region is spatially defined
in terms of a bivariate Gaussian distribution. Moreover, in the interest of
simplicity, we consider only cases where the similarity measure concerns a
single feature. In what follows, we present
two measures to quantify the similarity of two regions and discuss the merits
of each. Subsequently, we argue in favor of one of the two measures, and
present an algorithm to find similar regions according to that measure.
Finally, we describe some aggregate observations from employing
the two measures on our dataset.

\subsection{Similarity Measures}

We start by defining two similarity measures that have a natural interpretation
in our setting. The first one, \condsim, quantifies how similar the venues of
two regions are on average, according to the corresponding models.
The second one, \jointsim, combines two qualities of the regions:
(i) how similar the venues of
two regions are, but also (ii) how many venues they contain, according to the
model. The rationale for this second measure is that one might
not have a prior idea for how big the two regions should be and at the
same time want to avoid identifying pairs of tiny regions that are
`spuriously' similar.

\subsubsection{Similarity measure \condsim}
\label{sec:condsim}

In this section, we define \condsim, our first region similarity
measure, which operates under the following settings.
We are given two model instances, $I_1$ and $I_2$, each corresponding
to a city in our data. As explained in earlier sections of the paper, each
model instance describes the distribution of venues in its respective city,
along with distributions over every features. Moreover, we are given two bivariate Gaussian
distributions, $G_1$ and $G_2$, each defining one geographic region in the
respective model $I_1$ and $I_2$.
To define the similarity between the two regions with respect to feature $X$
(e.g., $X = $ {\tt category}),
we first define a random procedure $\mathcal{P}(I, G, X)$.
Given a model instance $I$ and a region $G$,
random procedure $\mathcal{P}$ generates one value $X = x$ for feature $X$.
It is defined as follows.
\begin{definition}[$\mathcal{P}(I, G, X)$]
Perform the following steps.
\begin{itemize}
  \item Generate a location $l \sim G$.
  \item Generate a data point at location $l$ from model $I$ and record
  its value $X=x$ for feature $X$.
  \item Return value $x$.
\end{itemize}
\end{definition}
In plain terms,
random procedure $\mathcal{P}$ picks a random location in region
$G$ and then generates a value for feature $X$ according to model instance $I$,
{\bf conditional} on the location selected in region $G$.
Similarity measure \condsim{} then answers the question: if random procedure
$\mathcal{P}$ is applied on model $I_1$ and region $G_1$ on one hand,
and model $I_2$ and region $G_2$ on the other, what is the probability
that it produces the same value $X = x$ for feature $X$?
In other words: if we compare two random venues, one from each region $G_1$
and $G_2$, what is the probability according to the respective models that
they would have the same feature $X$?
The measure is formally defined below.

\begin{definition}[\condsim]
Let $x_1$ and $x_2$ be the values of feature $X$ generated by a single
invocation of $\mathcal{P}(I_1, G_1, X)$ and $\mathcal{P}(I_2, G_2, X)$,
respectively. Similarity \condsim\ is defined as the probability
that $x_1 = x_2$.
\begin{equation}
  \condsim (G_1, G_2; I_1, I_2, X)
    = p(x_1 = x_2 | \mathbf{P})
\end{equation}
\end{definition}

\noindent We proceed to provide an analytical expression for \condsim.
First, in the interest of simplicity,
we fix model instances $I_1$, $I_2$ and feature $X$
and write $\condsim(G_1, G_2) = \condsim(G_1, G_2; I_1, I_2, X)$.
Let us also write $\gamma_i(x |l)$ to denote
the conditional probability that model $I_i$ ($i \in \{1, 2\})$,
generates a data point at location $l$ with feature $X$ taking value $x$
$$\gamma_i(x | l) = p(X = x | loc = l; I_i),$$
which can be expanded to
\begin{eqnarray*}
\gamma_i(x | l) = & p(X = x | loc = l; I_i) \\
                 = & \frac{p(X = x, loc = l | I_i)}{p(loc = l | I_i)} \\
    = & \frac{\sum_{z = 1..k}p(X = x, loc = l, topic = z | I_i)}
                {\sum_{z = 1..k}p(loc = l, topic = z | I_i)} \\
    = & \frac{\sum_{z = 1..k}{\mathcal{N}_z(l)\beta_z(v)\theta_z}}
                {\sum_{z = 1..k}{\mathcal{N}_z(l)\theta_z}},
\end{eqnarray*}
where $\mathcal{N}_z$ denotes the Gaussian probability density function for the
Gaussian distribution associated with the $z$-th topic, $z = 1..k$,
of model $I_i$.
Finally, for locations $l_1$ and $l_2$, let us write $g(l_1, l_2)$ for
the inner-product function
\begin{equation}
g(l_1, l_2) = \sum_{x \in Dom(X)} \gamma_1(x|l_1) \gamma_2(x|l_2),
\label{eq:g-inner}
\end{equation}
where $Dom(X)$ is the set of possible values for feature $X$.
With the notational conventions above, we now provide
an analytical expression for $\condsim$.
\begin{equation}
\condsim (G_1, G_2) =
  \int_{l_1, l_2} \mathcal{N}_1(l_1)\mathcal{N}_2(l_2) g(l_1, l_2) dl_1 dl_2
\label{eq:computecondsim}
\end{equation}
Note that, in equation~\eqref{eq:computecondsim}, $\mathcal{N}_1$ and
$\mathcal{N}_2$ denote the probability density functions for Gaussians
$G_1$ and $G_2$, respectively. In practice, we approximate the
integral of equation~\eqref{eq:computecondsim} by taking a discrete sum
over two $100\times 100$ grids of locations that cover the areas of the
corresponding cities.


\begin{figure}
\centering
\includegraphics[width=0.95\columnwidth]{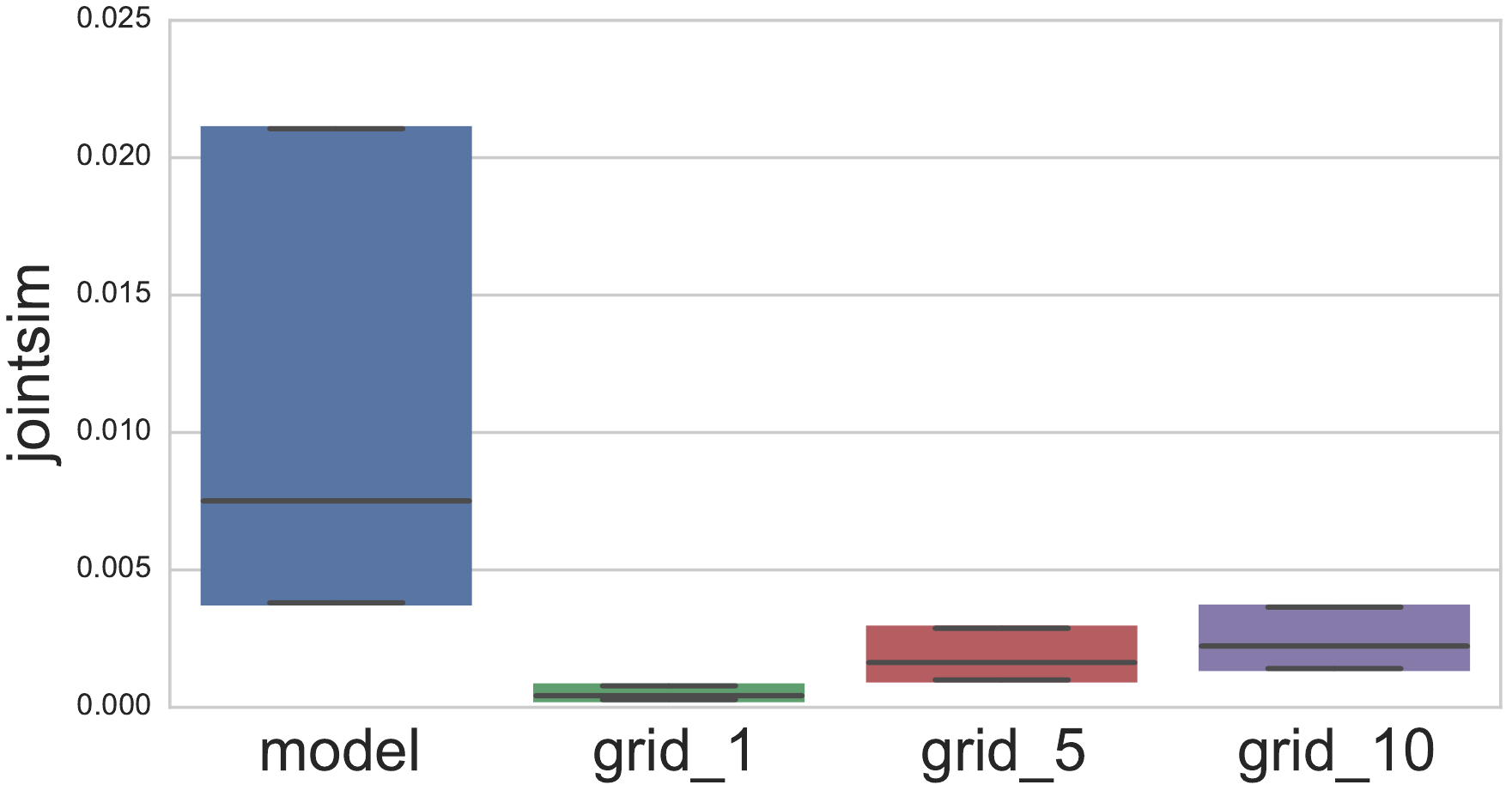}
\caption{The \jointsim\ values for the best-matching regions returned by
{\tt GeoExplore} across all
pairs of cities, for different sets of base-regions ({\bf model} and
{\bf grid-$\alpha$}).}
\label{fig:jointsim}
\end{figure}

\begin{figure}
\centering
\includegraphics[width=0.95\columnwidth]{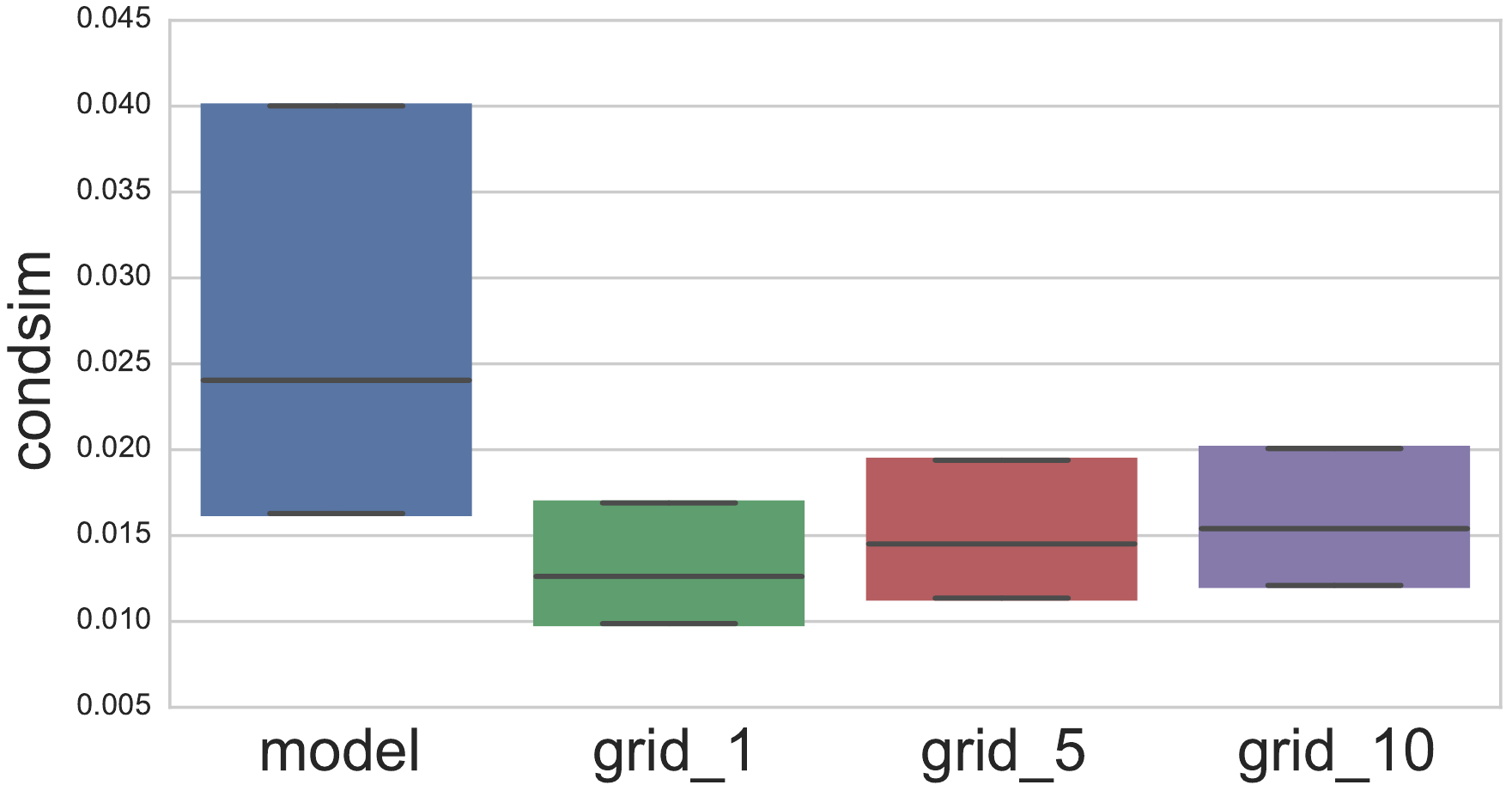}
\caption{The \condsim\ values for the best-matching regions across all
pairs of cities, for different sets of candidate regions ({\bf model} and
{\bf grid-$\alpha$}).}
\label{fig:condsim}
\end{figure}

\subsubsection{Similarity measure \jointsim}

In this section, we define \jointsim, our second region similarity
measure, which operates under the following settings,
similar to that of Section~\ref{sec:condsim}.
Namely, we are given two model instances, $I_1$ and $I_2$,
as well as two gaussians $G_1$ and $G_2$, each defining one region. We define
the similarity between the two regions with respect to feature $X$ on top of
a random procedure $\mathcal{R}(I, G, X)$, that generates a pair of values
$(x, s)$, with $x \in Dom(X)$ and $s\in \mathbb{R}^{+}$.
The random procedure $\mathcal{R}$ is defined as follows.
\begin{definition}[$\mathcal{R}(I, G, X)$]
Perform the following steps.
\begin{itemize}
  \item Generate a data point $d \sim I$, with location $loc = l$ and feature
  $X$ value $X = x$.
  \item Let $s = \mathcal{N}(l)$ be the probability density of $G$ at
  location $l$.
  \item Return the pair $(x, s)$.
\end{itemize}
\end{definition}
In plain terms, random procedure $\mathcal{R}(I, G, X)$ picks a random
data point from model instance $I$, and associates it with (1) its value
$X = x$ for feature $X$ and (2) the density of region-defining gaussian $G$.
Similarity measure \jointsim\ then answers the question: `if random
procedure $\mathcal{R}$ is applied on model $I_1$ and region $G_1$,
on one hand, and model $I_2$ and region $G_2$, on the other, with respective
output pairs $(x_1, s_1)$ and $(x_2, s_2)$, what is the
expected value of the expression
$$s_1 \cdot s_2 \cdot \delta_{x_1 x_2},$$
over possible invocations of procedure $\mathcal{R}$'?
In the expression above, $\delta_{x_1 x_2}$ is the Kronecker delta --
equal to one ($1$) only when $x_1 = x_2$ and zero ($0$) otherwise.
In other words, \jointsim\ combines the answer of the following two
questions: if we consider two random venues, one from each model $I_1$
and $I_2$, then (1) what is the probability that they have the same value for feature $X$,
and (2) if they do, how much are their locations covered by regions $G_1$ and
$G_2$? The measure is formally defined below:
\begin{definition}[\jointsim]
Let $(x_1, s_1)$ and $(x_2, s_2)$ be the output of a single
invocation of random procedure $\mathcal{R}(I_1, G_1, X)$ and
$\mathcal{R}(I_2, G_2, X)$, respectively. Similarity \jointsim\ is defined as
the expected value
\begin{equation}
\jointsim(G_1, G_2; I_1, I_2, X) =
  E_{\mathcal{R}}[s_1 \cdot s_2 \cdot \delta_{x_1 x_2}]
\end{equation}
\end{definition}

We now proceed to provide an analytical expression for \jointsim.
Following a similar derivation process as in Section~\ref{sec:condsim},
let us write $\psi_i(x, l)$ to denote the {\bf joint} probability
that model $I_i$ ($i \in \{1, 2\}$), generates a data point with feature value $X = x$
and location $l$,
$$\psi_i(x | l) = p(X = x, loc = l | I_i),$$
with
\begin{eqnarray*}
\psi_i(x | l) = & p(X = x , loc = l; I_i) \\
                 = & p(X = x, loc = l | I_i) \\
    = & \sum_{z = 1..k}p(X = x, loc = l, topic = z | I_i) \\
    = & \sum_{z = 1..k}{\mathcal{N}_z(l)\beta_z(v)\theta_z}
\end{eqnarray*}
Moreover, for locations $l_1$ and $l_2$, let us write $h(l_1, l_2)$ for
the inner-product function
\begin{equation}
h(l_1, l_2) = \sum_{x \in Dom(X)} \psi_1(x, l_1) \psi_2(x, l_2).
\label{eq:h-inner}
\end{equation}
With the notational conventions above, we now provide
an analytical expression for $\jointsim$.
\begin{equation}
\jointsim (G_1, G_2) =
  \int_{l_1, l_2} \mathcal{N}_1(l_1)\mathcal{N}_2(l_2) h(l_1, l_2) dl_1 dl_2
\label{eq:computejointsim}
\end{equation}
As in Section~\ref{sec:condsim}, $\mathcal{N}_1$ and
$\mathcal{N}_2$ in equation~\eqref{eq:computecondsim}, denote the probability
density functions of Gaussians
$G_1$ and $G_2$, respectively. Moreover, we approximate the
integral of equation~\eqref{eq:computecondsim} with a discrete sum approximation
over a $100\times 100$ grid on each model.

\subsubsection{Discussion}

Having defined the two similarity measures, let us consider
the corresponding maximization problems, first for \condsim{}.
\begin{problem}
Consider two models $I_1$, $I_2$, and feature $X$. Identify two bivariate
Gaussians $G_1$, $G_2$, such that $\condsim(G_1, G_2)$ is maximized.
\label{problem:condmax}
\end{problem}
One issue with problem~\ref{problem:condmax} is that it is prone to
return spurious regions, i.e., pairs of regions that are very similar,
but that cover too small probability mass of the respective models.
This can be seen from equations~\eqref{eq:g-inner} and~\eqref{eq:computecondsim}
that define \condsim:
they take into account the similarity of data points (venues) in terms of
feature $X$ at locations within the two regions, but they do not take into
account the probability mass that the respective models assign to those
regions.
To provide a contrived but illuminating example, one can consider
the case where $G_1$ and $G_2$ cover very small areas with few venues
that are identical w.r.t. feature $X$. These two regions are thus
associated with higher \condsim\ values than larger areas whose venues
are similar, but not identical w.r.t. feature $X$.
Therefore, if one were to find optimal solutions for
Problem~\ref{problem:condmax}, they would have to consider such spurious
cases as optimal solutions to the problem. Since we have no good way
to deal with this issue, we are forced to impose constraints
on the possible candidate gaussians $G_1$ and $G_2$
that comprise the solution pair. Specifically, we consider
Problem~\ref{problem:modified-cond} as a specific case of
Problem~\ref{problem:condmax}, in which the candidate solution pairs are
provided as input.
\begin{problem}
Consider two models $I_1$, $I_2$, and feature $X$.
Consider also two collections of Gaussians
$\mathbb{G}_1$, $\mathbb{G}_2$.
Identify two bivariate
Gaussians $G_1 \in \mathbb{G}_1$, $G_2 \in \mathbb{G}_2$,
such that $\condsim(G_1, G_2)$ is maximized.
\label{problem:modified-cond}
\end{problem}

Let us now consider the corresponding optimization problem for \jointsim{}.
\begin{problem}
Consider two models $I_1$, $I_2$, and feature $X$. Identify two bivariate
Gaussians $G_1$, $G_2$, such that $\jointsim(G_1, G_2)$ is maximized.
\label{problem:jointmax}
\end{problem}
Problem~\ref{problem:jointmax} has two intuitive properties:
\squishlist
\item all other things being equal, it favors regions $G_1$ and $G_2$ that
cover areas with high probability mass according to
models $I_1$, $I_2$,
\item all other things being equal, it favors regions $G_1$ and $G_2$ that
cover areas where data points are associated with similar feature
distributions.
\squishend
To put in plain terms, Problem~\ref{problem:jointmax} favors regions that
correspond to areas of {\bf many} {\it and} {\bf similar} data points.
This is seen
also in equations~\eqref{eq:h-inner} and \eqref{eq:computejointsim}. Indeed,
they take into account the similarity of data points (venues) in terms of
feature $X$ at locations within the two regions, but they also take into
account the probability mass assigned to those regions by models $I_1$ and
$I_2$. Therefore,
Problem~\ref{problem:jointmax}
does not suffer from the spurious solutions problem that affected
Problem~\ref{problem:condmax}. This makes Problem~\ref{problem:jointmax}
appropriate to consider in cases when one does not have prior restrictions
or preferences for the candidate regions that would comprise an optimal
solution (as in the formulation of Problem~\ref{problem:modified-cond})
and at the same time would like to avoid spurious solutions.

\subsection{Best-First Search for Joint Similarity}

To the best of our knowledge, the problem of identifying
similar regions across cities has not been defined formally
before within a probabilistic framework.
In this section, we also propose the first algorithm, \geoxpl, to approach
Problem~\ref{problem:jointmax}.
Algorithm \geoxpl\ follows a typical {\it best-first} exploration scheme and
comprises of the following two phases.
Its first phase consists of one step:
it begins with a candidate collection of regions $\mathbb{G}_1$ and
$\mathbb{G}_2$ for each side (let us call them `{\bf base regions}') and
evaluates all pairwise similarities $\jointsim(G_1, G_2)$,
for $G_1\in\mathbb{G}_1$, $G_2\in\mathbb{G}_2$.
Its second phase consists of the remaining steps:
it explores the possibility
to improve the currently best \jointsim\ measure by combining previously
considered regions. This is motivated by the fact that
Problem~\ref{problem:jointmax} favors regions of larger probability mass, and
therefore combined regions might yield better \jointsim\ values.

Pseudocode for \geoxpl\ is shown in Algorithm~\ref{algo:explore}.
It repeats
a three-steps {\it Retrieve - Update - Expand} procedure for each step.
During {\it Retrieve}, the algorithm retrieves the next candidate solution
for Problem~\ref{problem:jointmax}. Each candidate solution comes in the
form of a triplet; two Gaussians and their \jointsim\ score.
During {\it Update}, the algorithm updates the score of the best-matching pair,
if a better pair has just been retrieved.
Finally, during {\it Expand}, the algorithm expands the latest retrieved
Gaussians to form Gaussians from each side, and thus new candidate
solutions for Problem~\ref{problem:jointmax}. Subroutine {\bf expand(G, I)}
operates as follows:
\squishlist
\item When $G$ is not specified (i.e., $G = \text{NULL}$
in Algorithm~\ref{algo:explore}), then {\bf expand} simply returns the set of
base regions $\mathbb{G}$. This case occurs during the first expansions only.
Moreover, each base region $G_i \in \mathbb{G}$ is associated with positive
weight $w_i$, either specified as input, or set to $1 / \|\mathbb{G}\|$ by
default.
\item When $G = G_i$ for some $G_i \in \mathbb{G}$, then {\bf expand}
returns the set of Gaussians
$\{G_i\} \bigcup \{G_i \cup G_j; G_j \in\mathbb{G}, j\neq i\}$,
where $G_i \cup G_j$ is defined as the best Gaussian-fit to the mixture model
determined by $[G_i$, $G_j]$, with respective proportions
$(w_i$, $w_j)$. The intuition for this step is that we expand
the best-performing pair of Gaussians by combining them with other base
Gaussians.
\item In a recursive fashion, when $G = G_i\cup G_{i'}\ldots G_{i''}$, then
{\bf expand} returns the set of Gaussians
$\{G\} \bigcup \{G \cup G_j; G_j \in\mathbb{G}, j\neq i, i', \ldots, i''\}$
each defined  as the best Gaussian-fit to the mixture model
determined by $[G$, $G_j]$, with respective proportions
$(w_i + w_{i'} + \ldots + w_{i''}$, $w_j)$.
\squishend
Note that in practice, to prevent the algorithm from
exploring the combinatorially large space, we terminate \geoxpl\ after
a number $R$ of {\bf while} loops.

\begin{algorithm}[t]
\caption{\geoxpl}
\label{algo:explore}
\begin{algorithmic}
\STATE {\bf Input}: models $I_1$ and $I_2$, base regions $\mathbb{G}_1$,
$\mathbb{G}_2$
\STATE {\bf Output}: Best Pair $G_1$, $G_2$
\STATE
\STATE \# INITIALIZE
\STATE $\text{BestG}_1 = \text{NULL}$, $\text{BestG}_2 = \text{NULL}$, \
    BestScore = 0
\STATE H = MaxHeap()
\STATE \# Initialize max-heap with empty solution, zero score
\STATE Push($\text{BestG}_1$, $\text{BestG}_2$, BestScore) to H
\STATE
\WHILE {H is Not Empty}
\STATE \# RETRIEVE top solution in max-heap
\STATE Pop ($\text{G}_1$, $\text{G}_2$, Score) from H
\STATE \# UPDATE best solution
\IF{Score $>$ BestScore}
\STATE $\text{BestG}_1 = {\text G}_1$, $\text{BestG}_2 = {\text G}_2$, \
    BestScore = Score
\ENDIF
\STATE \# EXPAND retrieved solution
\FOR{$\text{G}_a$, $\text{G}_b$ in {\bf expand($\text{G}_1$)}, \
{\bf expand($\text{G}_2$)}  $\neq \text{G}_1$, $\text{G}_2$}
\STATE Score = $\jointsim(G_a, G_b | I_1, I_2)$
\STATE Push ($\text{G}_a$, $\text{G}_b$, Score) to H
\ENDFOR
\ENDWHILE
\STATE \textbf{return} best\_pair
\end{algorithmic}
\end{algorithm}

\subsection{Empirical performance}

We employed \geoxpl\ with $R = 5$ expansions
on all pairs of cities in our dataset
(Section~\ref{sec:dataset}) and report the \jointsim\ values
returned for different {base region} collections.
Specifically, we experimented with the following collections of base
regions:
\begin{description}
  \item[Model] We simply used as collections $\mathbb{G}_1$, $\mathbb{G}_2$
  the Gaussians associated with the respective
  topics in the input model $I_1$, $I_2$, and assigned to each
  Gaussian a weight equal to the respective $\theta$ parameter value
  found in the model.
  \item[Grid-$a$] We used as collections $\mathbb{G}_1$, $\mathbb{G}_2$
  Gaussians that covered in a grid-like fashion the respective cities,
  each with size equal to $1/a$ the size of the median size of Gaussians
  found in model $I_1$, $I_2$.
\end{description}
The results are shown in Figure~\ref{fig:jointsim}. We observe that
using the model Gaussians as our base regions leads to better performance
compared with the grid baselines.

Finally, in Figure~\ref{fig:condsim}, we report the values obtained
for Problem~\ref{problem:modified-cond} by a straightforward
all-pairs algorithm. The collections of Gaussians $\mathbb{G}_1$ and
$\mathbb{G}_2$ provided as input to the problem are defined in the
same way as the base regions of Figure~\ref{fig:jointsim} above.
Again, we observe that we obtain the best performance in terms of
\condsim\ when the candidate regions coincide with the model Gaussians.


\section{Comparison with previous approaches}
\label{sec:other_approaches}

In this section, we compare empirically our approach with previous work.
Ideally, our comparison would be with works that address the same task as
this paper -- i.e., model the distribution of venues across a city.
In such a case, we would have a natural and direct measure of comparison, namely
the predictive performance of each model in terms of log-likelihood.
However, to the best of our knowledge, no such work is readily available.
Therefore, our comparison is with previous work that
addressed slightly different different tasks. Nevertheless, the comparison
serves as a `sanity check' for our approach and allows us to
understand better the proposed technique.

We compare with two methods which provide publicly available results,
namely Livehoods\footnote{\url{http://livehoods.org/maps}} and
Hoodsquare\footnote{\url{http://pizza.cl.cam.ac.uk/hoodsquare}}.
The obtained results concern three large US cities that also appear in our 
dataset.
Both methods use Foursquare data and output a geographical clustering
of venues within a city, with each such cluster defining one region on
the map of the city.  
In the case of \livehoods{}~\cite{Livehoods12}, the clusters are obtained through spectral
clustering on a nearest-neighbor graph over venues, where the edge weights
quantify the volume of visitors who check-in at both adjacent venues.
In the case of \hoodsquare{}~\cite{Hoodsquare13}, clusters are obtained by employing the OPTICS
clustering algorithm on venues, using a number of venue features
(location, category, peak time of activity during the day, and a binary
touristic indicator).

\begin{figure*}[tb]
  \centering
    \subfigure[Hoodsquare]{
      \includegraphics[width=0.3\linewidth]
      {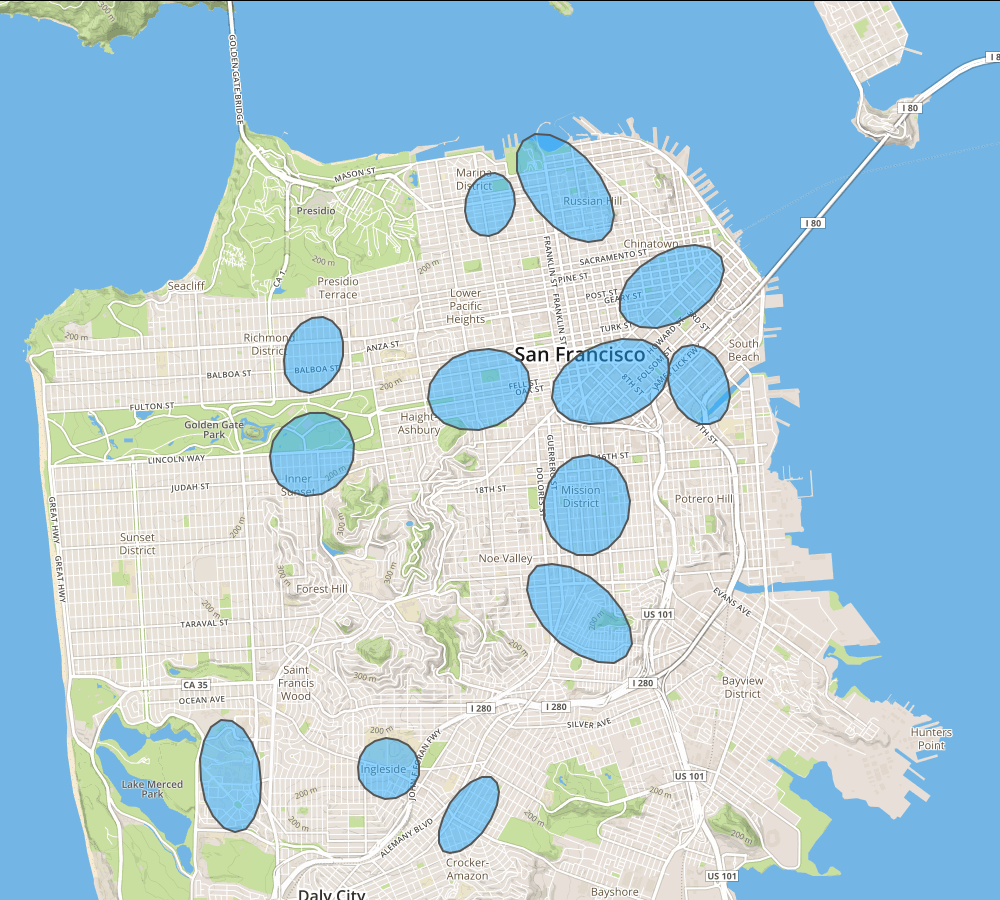}
    }
    \subfigure[Our method]{
      \includegraphics[width=0.3\linewidth]
      {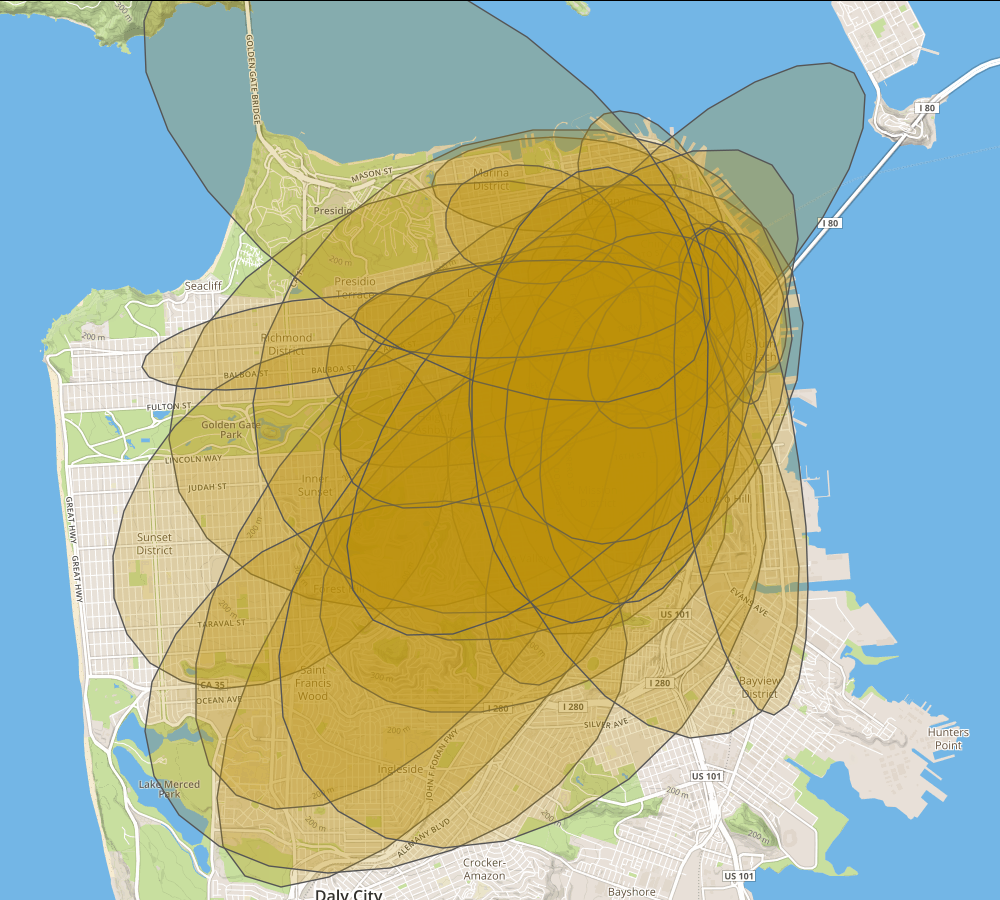}
    }
    \subfigure[Livehoods]{
      \includegraphics[width=0.3\linewidth]
      {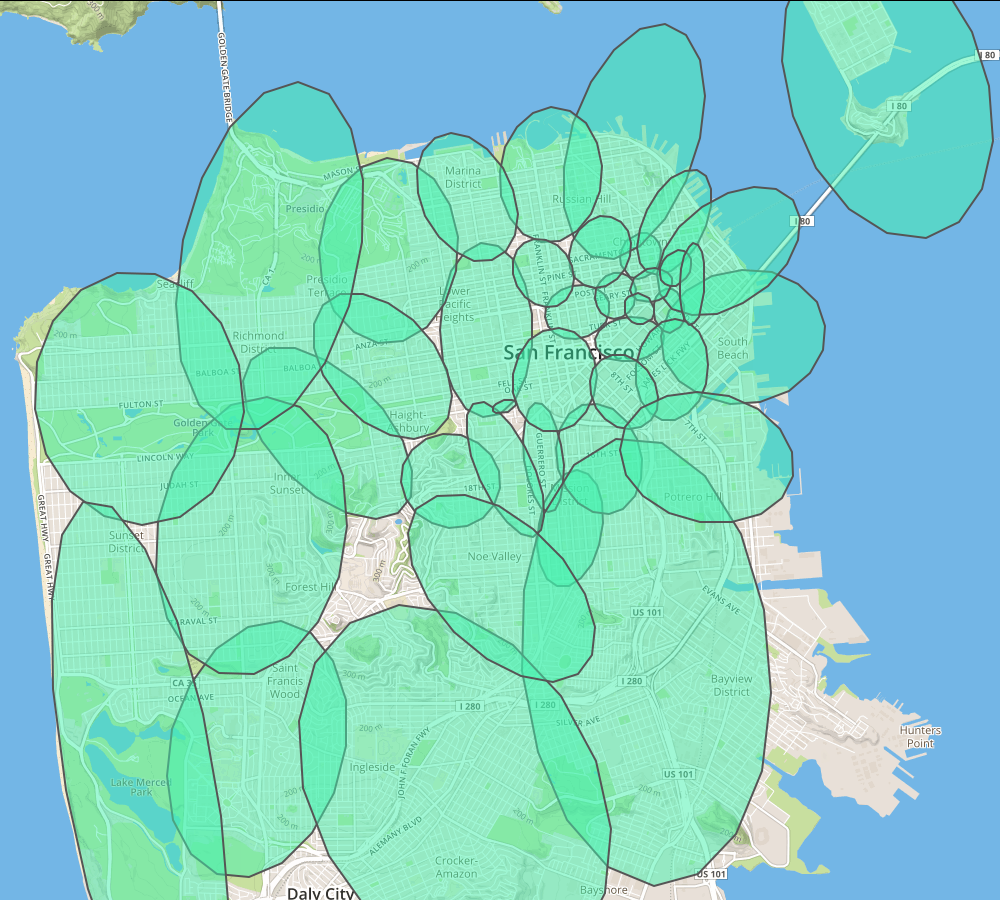}
    }
  \caption{In San Francisco, the middle panel shows the regions we
      obtained with our probabilistic model. On the left, Hoodsquare
      regions are extracted from their website and transformed into
      Gaussians. On the right is the same picture for Livehoods. All methods show that
      the city activity occurs mainly downtown but it also highlights
      differences between approaches. For instance, although Livehoods
      exhibits some overlap, it is only due to the Gaussian conversion whereas
      we do not restrain venues to serve a single function in a single
      region by construction, but only let that happens if the data
      support it.
\label{fig:livehoods}}
\end{figure*}

To compare, we perform the following procedure.
First, we obtain the clusters returned by each method.
We interpret each of those clusters as a region that belongs to one {\it topic},
in the sense that we've been using the term in the context of our model.
To map them to our setting, we approximate the shape of each region with
the smallest bivariate Gaussian, so that the entire region is enclosed within
two standard deviations of the Gaussian in each direction.
(see Figure \ref{fig:livehoods} for the visual results of this approximation in San Francisco)
In this way, we obtain a number $k$ of Gaussians from each method.
We then train an instance of our model on our data, using the same number
$k$ of topics, and keeping the Gaussians associated with each topic fixed
to the Gaussians extracted with the aforementioned steps.
As in previous sections of the paper, we hold out 20\% of the data as test
dataset, on which we evaluate the log-likelihood of each learnt model instance.

The log-likelihood achieved by the different models is shown in the first
row of Table \ref{tab:lvhq}. As one can see there, the results based on our
model perform better in predicting the test dataset.
This is not surprising, since our approach optimizes predictive accuracy
directly. Nevertheless, the results provide evidence that our approach works
reasonably well for the task it was designed to address.

To further quantify the differences between the three approaches, we report
additional quantities from the learned models, described below. Essentially
those quantities capture how distinct the identified regions of each model
are in terms of the associated features.

\spara{Mean Feature Entropy} We consider each categorical feature
separately and--for each topic region in the respective model of the
three approaches--we measure the entropy of the respective multinomial
distribution $\beta$. Intuitively, we would like the regions that
constitute our model instances to capture the variance of the various features
across the city. Therefore, we would like the $\beta$ distributions of the
model instances to have lower entropy (i.e., be farther from uniform).
Table \ref{tab:lvhq} reports the mean entropy of $\beta$ across regions for each
categorical feature and for each of the methods in the three US cities.
The relevant lines in the table are the ones labelled `mean [feature]
entropy'. 
As one can confirm, in the majority of cases the model instance based on
our method returns $\beta$ distributions of lower entropy.

\spara{Jensen--Shannon Divergence from City Average} Another way to quantify
the distinctiveness of the various regions is to measure the distance of
the $\beta$ distribution of each feature and topic in the model from
the average distribution $\mu$ for the same feature across the city.
One principled approach to quantify this difference is to use 
Jensen--Shannon divergence, ${\rm JSD}(\beta, \mu)$ a symmetrized
version of the Kullback--Leibler divergence $\mathrm{KL}(P \parallel Q)$.
It is defined with the following formula.
$${\rm JSD}(\beta, \mu)= 0.5\cdot \mathrm{KL}(\beta \parallel
(\beta+\mu)/2)+0.5\cdot \mathrm{KL}(\mu \parallel (\beta+\mu)/2))$$
Intuitively, it is desirable for $\beta$ distributions of different topics
to differ from average city behavior as captured by $\mu$ distributions.
Table \ref{tab:lvhq} reports the average Jensen--Shannon divergence across the
topics for each categorical feature, city, and method.
The relevant lines in the table are the ones labelled `[feature] JSD from 
city'. Again, in the majority of cases, the model instance
based on our method returns $\beta$ distributions that differ more
from city average distribution than the methods we compare with.

\mpara{} To summarize our findings from Table \ref{tab:lvhq},
our model has better predictive performance, while generally identifying 
topic regions that are more distinct with each other and further from
average, despite their high overlap. The results thus provide evidence that
our approach discovers regions with desirable properties.

\begin{table*}[tb]
\setlength{\tabcolsep}{1.4pt}
\centering
\footnotesize
\caption{Comparison in San Francisco, New York and Seattle between our model, Livehoods (LH) and
Hoodsquare (HS, which has no neighborhoods available in Seattle). The last
abbreviation, JSD, stands for the average Jensen--Shannon divergence between
regions and city-wide distributions of the four features we consider: category,
users, dayOfWeek and timeOfDay. Each group of two adjacent columns is a
comparison betwen a competing method and our model. The arrow after the name of each
measure indicates whether better values are higher or lower.
\label{tab:lvhq}}
\begin{tabular}{lcrr|cc|cc|cc|cc}
\toprule
& & \multicolumn{4}{c}{San Francisco} & \multicolumn{4}{c}{New York} & \multicolumn{2}{c}{Seattle} \\
{} & {} & HS & Us & LH & Us & Hoodsquare & Us & Livehoods & Us & Livehoods & Us \\
\midrule                                                                                                                                                                                               
likelihood per venue       & $\nearrow$ & \cwin{-202.9}  & \cwin{-198.3}  & \cwin{-197.0}  & \cwin{-195.4}  & \cwin{-343.9}  & \cwin{-277.9}  & \cwin{-271.1} & \cwin{-268.2} & \cwin{-112.6}  & \cwin{-112.2}  \\
\midrule                                                                                                                                                                                                           
mean category entropy      & $\searrow$ & \cwin{4.961}   & \cwin{4.632}   & \cwin{4.802}   & \cwin{4.776}   & \cwin{4.886}   & \cwin{4.714}   & \cwin{4.851}  & \cwin{4.716}  & \cloose{5.178} & \cloose{5.307} \\
category JSD from city  & $\nearrow$ & \cwin{0.053}   & \cwin{0.092}   & 0.065          & 0.065          & \cwin{0.054}   & \cwin{0.085}   & \cwin{0.055}  & \cwin{0.070}  & \cloose{0.016} & \cloose{0.001} \\
\midrule                                                                                                                                                                                                           
mean dayOfWeek entropy     & $\searrow$ & \cwin{1.908}   & \cwin{1.894}   & \cloose{1.892} & \cloose{1.896} & 1.923          & 1.923          & 1.917         & 1.917         & \cwin{1.896}   & \cwin{1.864}   \\
dayOfWeek JSD from city & $\nearrow$ & \cwin{0.007}   & \cwin{0.013}   & 0.012          & 0.012          & 0.005          & 0.005          & 0.006         & 0.006         & \cwin{0.009}   & \cwin{0.016}   \\
\midrule                                                                                                                                                                                                           
mean timeOfDay entropy     & $\searrow$ & \cwin{1.488}   & \cwin{1.440}   & \cwin{1.428}   & \cwin{1.414}   & \cloose{1.522} & \cloose{1.542} & 1.535         & 1.535         & \cwin{1.477}   & \cwin{1.395}   \\
timeOfDay JSD from city & $\nearrow$ & \cwin{0.018}   & \cwin{0.034}   & \cwin{0.031}   & \cwin{0.037}   & \cwin{0.018}   & \cwin{0.025}   & \cwin{0.021}  & \cwin{0.024}  & \cwin{0.022}   & \cwin{0.041}   \\
\midrule                                                                                                                                                                                                           
mean users entropy          & $\searrow$ & \cloose{5.610} & \cloose{5.617} & \cwin{5.522}   & \cwin{5.305}   & \cwin{5.929}   & \cwin{5.609}   & \cwin{5.231}  & \cwin{4.975}  & \cwin{5.076}   & \cwin{5.058}   \\
user JSD from city      & $\nearrow$ & \cwin{0.160}   & \cwin{0.170}   & \cwin{0.175}   & \cwin{0.201}   & \cwin{0.092}   & \cwin{0.162}   & \cwin{0.198}  & \cwin{0.224}  & \cwin{0.187}   & \cwin{0.192}   \\
\bottomrule
\end{tabular}
\end{table*}

\section{Related Work}
\label{sec:related}

Urban Computing is an active area of research, partially due to the
increasing volume of digital data related to human activity and the potential
to use such data to improve life in cities. Below, we discuss related works
that either address a similar task (finding geographical structure in 
city activity) or use a similar approach to ours to address different
tasks (Section~\ref{sec:related_city_structure}).
To the best of our knowledge, we are the first to employ a fully probabilistic
approach to this task -- and thus discuss further the concept of sparse
modeling (Section~\ref{sec:related_sparse}).
Moreover, we discuss briefly other tasks pertaining to Urban
Computing that are more loosely related to our
work (Section~\ref{sec:related_other}).

\subsection{Finding structure in urban activity from digital traces}
\label{sec:related_city_structure}

Finding cohesive geographical regions within cities has been attempted using
a variety of data sources, such as cellphone activity~\cite{CDRLandUse12},
geotagged tweets~\cite{spectralLandUseTwitter14},
social interactions~\cite{Hipp2012128},
types of buildings~\cite{CityComposition13},
or public transport and taxi trajectories~\cite{zoneTrajectories15,taxiChina15}.

In that context, Location Based Social Networks ({\it LBSNs})
have also proven a rich source of data and utilized by recent works.
For instance, \cite{Livehoods12} collects checkins and build a $m$-nearest
spatial neighbors graph of venues, with edges weighted by the cosine similarity
of both venue's user distribution. The regions are the spectral clusters
of this graph.
Using similar data, \cite{Hoodsquare13} describes venues by category, peak
time activity and a binary touristic indicator. Venues are clustered in
hotspots along all these dimensions by the \textsc{OPTICS} algorithm. The city
is divided into a grid, with cells described by their hotspot density for each
feature. Finally, similar cells are iteratively clustered into regions.
Like us, \cite{FunctionalArea15} considers venues to be essential in defining
regions. The city is divided into a grid of cells with the goal of
assigning each cell a category label in a way that is as specific as possible while
being locally homogeneous. This is done through a bottom-up clustering which
greedily merge neighboring cells to improve a cost function formalizing this
trade off.

While these results are evaluated with user interviews and built upon well
known algorithmic techniques, they rely on ad-hoc modeling decisions
(such graph construction and grid granularity) that do not
derive directly from the data, and thus leave us at a loss when we are asked to
support the statistical significance of the obtained results.
Furthermore, because the clustering is not guided simultaneously by all the
available data features, such as time and aspects other than venue category,
important information might be going amiss in those approaches.


On the other hand, there are works that take a probabilistic
approach, although their aim is different than ours.
For instance, \cite{cranshaw2010seeing} assigns venues to a grid and
runs Latent Dirichlet Allocation (LDA) on their categories.
However it does not output explicit regions, and the grid is coarse
approximation for using spatial information. Instead, 
\cite{GeoTopicYin11} fixes a number $K$ of localized topics to be discovered,
as well as a set of $N$ Gaussian spatial regions. Each region has a topic
distribution and each topic is a multinomial distribution over all possible
Flickr photos tags. Relaxing several assumptions, notably the Gaussian shape,
\cite{NonGaussianTopicKling14} extends Hierachical Dirichlet Process to
spatial data, giving rise to an almost fully non-parametric model
(as the number of regions still needs to be set manually).
One extension of such methods is to take users features into account, in order
to provide recommendations~\cite{GeoTopicKurashima2013}.
While such methods bear similarity with ours, the domain of application
presented by their authors forbids direct comparisons.

Closer to the task of finding regions, \cite{UrbanStory12} performs LDA
on checkins in New York.
The five resulting topics are called urban activities, and venues are clustered
by their topic proportion across time. Contrary to us, this clustering does not
produce clearly defined regions as it is done as a post processing step.
Indeed, their LDA model does not incorporate a spatial dimension.
Moving from checkins to a dataset of 8 millions Flickr geotagged photos,
\cite{Kafsi2015}
probabilistically assigns tags to one of the three levels of a spatial
hierarchy, where each node is associated with a multinomial distribution over
tags. This allow to find the most descriptive tags for a given
regions and thus characterize it. Like in our work, a measure of
similarity between regions is defined.

\subsection{Sparse topic model}
\label{sec:related_sparse}

We now present related applications of the Sparse additive generative model on
spatial data. 
The original SAGE paper~\cite{eisenstein2011sparse} evaluates its model
effectiveness on the task of predicting the localization of twitter users by
learning not only topics about words but also about metadata (i.e. in which
region was the tweet written) and shows good accuracy. 
Later, a simplified version of it was used to find regions which exhibit 
geolocated idioms~\cite{eisenstein2015geographical}.
It is possible to better model user location by building a hierarchy of regions
\cite{nestedChinese13}.
Even though the sparsity of model is well suited to the sparsity of textual
data, note that methods which do not use topic modelling give competitive 
results in terms of accuracy~\cite{locatingtweet14,hyperlocalgeotagging15}.

Another task hindered by data sparsity and which benefits from modeling user
preferences is spatial item recommendation.
The interested reader will find
many examples exploiting LBSN in a recent survey~\cite{LBSNRecommendation15},
but here we give a taste of two approaches inspired by SAGE.
In both cases, topics are distributions over words and venues. Each
user is endowed with her own topic, and so to are each region.
\cite{SpatialTopicRecommendation13} uses SAGE to model user topics as a
variation from the overall global distribution.
To improve out of town recommendation, \cite{Wang2015a} assigns regions both
local and tourist topics. Learning such high number of parameters is made
possible by combining SAGE and a hierarchical model called spatial pyramid.

\subsection{Other urban computing problems}
\label{sec:related_other}

After finding regions in a city, a natural task is to compare them, within or
across cities. One might also look at different granularity to perform such
urban comparisons, whether points of interest or cities as a whole.
Finally, one might focus on users and model their preferences through mobility data.

\begin{description}[nosep,wide]
\item[Comparing regions]
We saw that one way to compare regions is to assign them
descriptive tags~\cite{Kafsi2015}. Others have looked at a more information
retrieval approach~\cite{SimilarRegion10},
while in~\cite{Thesis15}, authors collect data from
Foursquare and Flickr to associate venues with a features vector
summarizing their time activity, their surroundings, their category and
their popularity. They then define the
similarity between two regions as the Earth Mover Distance
between the two sets of venues feature vectors they contain. Finally they
devise a pruning and clustering strategy to perform efficient search.
\item[Finding points of interest]
Regions are only one subdivision of the city, another one are points of
interest, which are locations where the user activity show some specificities.
Geotagged photos can be mined to extract the semantics of such locations
\cite{Deng2009,Hotspots12,TagHotspot12,Feick2014,Hu2015}. As a
representative example, \cite{Rattenbury2009} compares various spatial
methods to discover small areas in San Francisco where one photo tag appears in
burst.
GPS trajectories provide useful information as well
\cite{TrajROI11,SemanticTrajectories14,TrajSurvey13}. For instance,
\cite{GPSStay10} extracts stay points from car GPS data and assess their
significance by how many visitors go there, how far they
traveled to reach them and how long they stay.
\item[Comparing cities] This has been done by comparing
the spatial distribution of human hotspots using call data~\cite{PhoneCityStructure14},
the call data profile themselves~\cite{ComparativeCity15},
the distribution of venues category at various scales~\cite{CitiesSimilarity13}, or by
building a network of city from urban residents mobility flow and computing
centrality measures~\cite{CityInfluence15}.
\item[Clustering users] With venues, users are the other side of the coin
of what defines a region in a city, and some works have mine their activities
to extract meaningful groups. For instance, \cite{ClusteringUsers13}
clusters users by the similarity of their venues category transition
probability matrix. Another approach is to consider users as document,
checkins as word and apply LDA, thus revealing cohesive
communities~\cite{UserCluster12} and describing people lifestyle~\cite{Lifestyle13}.
\end{description}

One can find other applications of
Urban Computing in related literature 
surveys \cite{UrbanComputingSurvey14,Tasse2014}.

\section{Conclusions}
\label{sec:conclusions}

In this work, we made use of a probabilistic model to reveal how
venues are distributed in cities in terms of several features.
As most habitants of a city do not visit most of the available venues, we
cope with the induced sparsity by adapting the sparse modeling approach of
\cite{eisenstein2011sparse} to data at hand.
Fitting our model to a large dataset of more than 11 million
checkins in 40 cities around the world, we show the insights provided by such
an unsupervised approach.

First, using the extracted model instances, we calculated
the probability distribution of a single feature conditioned
on the location in the city. This enabled us to construct a
heatmap of that feature to highlight what feature values are most likely
and distinctive at different locations within a city.

Secondly, we described a principled approach to quantify the importance
of different features within the trained models. Whereas all features
contribute, we discovered that the most defining feature for the components
uncovered by the model is the visitors of venues. This finding suggests
that further analysis of user behavior is a promising direction for
extracting additional insights.

Third, after focusing on the various regions of a single city,
we used the extracted model instances to find the most two similar
regions between two cities, a task which was previously attempted
with a more heuristic approach~\cite{Thesis15}.
This time we also benefit of the solid theoretical grounds of probabilistic
models to define principled measures of similarity and we describe a procedure
to greedily find two regions which maximize such measures.

Finally, we compare our approach with previous approaches that provide
similar output and show that our regions are both more consistent with the
data (in terms of predictive performance) and have more sharply defined
characteristics, meaning they are easier to distinguish from one another.

A review of recent related works in the Urban Computing field suggests that
whereas the area is active and that understanding urban activities is a worthy
endeavour which benefits from geotagged data, it can be pushed further by the
use of probabilistic models, as such models come with great
interpretative power.

Looking beyond this paper, there are further directions in which we
can improve our discovery process, providing additional interpretation along
the way:
\begin{itemize}
\item The first direction is to use a different evaluation process, one
  that would involve more closely users, since we show they are the main
  actor of the regions we discover. For instance, this could take the form of
  interviews. The purpose of this would be to identify
  and correct, if any, significant biases that are embodied in particular
  datasets (\fsq{} activity, in our case).
\item The model itself could
  also be extended, for instance by incorporating hierarchies of regions.
  This would both provide more structure to our results and allow us to
  apply our method to larger geographical area while keeping sparsity and
  runtime under control. Hierarchy is a well studied concept in both
  spatial data mining and topic modelling \cite{NestedTopics10} and thus we
  are confident this would be a feasible improvement. Another direction
  would be to incorporate additional features into the model (e.g.,
  continuous features)
\item Just as natural landscapes change with time, whether because
  of the day/night cycle or the passing of seasons, so do cities. It is not
  far fetched to imagine than coastal areas of Barcelona or San Francisco
  witness different patterns of activities in the winter than in the
  summer. Again following the time evolution of topics has been addressed in
  different settings \cite{DynamicTopic06}.
\item There is the opportunity to use the automatically extracted
  models to build advanced systems. One direction would be to
  involve users in a discovery-and-feedback process, where users
  would indicate regions they appreciate in their hometown and the
  system would help them plan a trip in a new city based on model instances
  trained for the two cities.
\end{itemize}



\begin{IEEEbiographynophoto}
{Emre Çelikten} is a graduate student at Aalto University.
Previously he has worked as a software engineer focusing on scalable web back-end
development. He received his Bachelor's degree from the Computer Engineering
department of İzmir Institute of Technology.
He is currently serving as research assistant at Aalto University.
His research interests include machine learning and mining social media data.
\end{IEEEbiographynophoto}
\begin{IEEEbiographynophoto}
{Géraud Le Falher} is a PhD student at Inria Lille.
He received his Master of Engineering from École Centrale de Nantes, France
and his Master's degree from the Computer Science department of Aalto University.
His research interests include mining urban data and learning with graph data,
especially in the context of signed graphs.
\end{IEEEbiographynophoto}
\begin{IEEEbiographynophoto}
{Michael Mathioudakis} is a
postdoctoral researcher at the Helsinki Institute for
Information Technology HIIT and Aalto University. He received his PhD
from the Department of Computer Science at the University of Toronto in 2013.
His research interests focus mostly on the analysis of user-generated
content on the Web, including the analysis of information diets on the
Web, urban informatics, and trend detection on social media.
\end{IEEEbiographynophoto}

\end{document}